\documentclass[12pt]{article}
\pdfoutput=1
\usepackage{cite}
\usepackage{enumerate}
\usepackage{ifpdf}
\usepackage[T1]{fontenc}
\usepackage[ansinew]{inputenc}
\usepackage{amsmath}
\usepackage{relsize}
\usepackage{amssymb}
\usepackage{tikz}
\usepackage{graphicx}
\usepackage{color}
\definecolor{darkblue}{cmyk}{0.9,0.9,0,0}
\definecolor{darkgreen}{rgb}{0,0.55,0}
\usepackage[colorlinks=true,linkcolor=darkblue,citecolor=darkblue,urlcolor=darkblue]{hyperref}
\usepackage{epsfig}
\usepackage{epstopdf}
\usepackage{graphicx}
\usepackage{mathtools}
\usepackage[marginpar=2cm]{geometry}
\DeclarePairedDelimiter\floor{\lfloor}{\rfloor}

\makeatletter
\long\def\@makecaption#1#2{
  \vskip\abovecaptionskip
  \sbox\@tempboxa{{\captionfonts #1: #2}}
  \ifdim \wd\@tempboxa >\hsize
    {\captionfonts #1: #2\par}
  \else
    \hbox to\hsize{\hfil\box\@tempboxa\hfil}
  \fi
  \vskip\belowcaptionskip}
\makeatother

\newcommand{\beq}{\begin{equation}}
\newcommand{\eeq}{\end{equation}}
\newcommand{\beqy} {\begin{eqnarray}}
\newcommand{\eeqy} {\end{eqnarray}}
\newcommand{\bsmat}{\begin{smallmatrix}}
\newcommand{\esmat}{\end{smallmatrix}}
\newcommand{\bmat}{\begin{matrix}}
\newcommand{\emat}{\end{matrix}}

\newcommand{\be}{\begin{equation}}
\newcommand{\ee}{\end{equation}}
\newcommand{\bea}{\begin{align}}
\newcommand{\eea}{\end{align}}
\newcommand{\nn}{\nonumber}

\def\({\left(}
\def\){\right)}
\def\[{\left[}
\def\]{\right]}

\def\<{\langle}
\def\>{\rangle}

\usepackage{varioref}
\usepackage{makeidx}
\makeindex

\usepackage[english]{babel}
\usepackage{caption}
\usepackage{subfig}

\usepackage{tabularx}
\newcommand{\Mod}[1]{\ (\mathrm{mod}\ #1)}

\begin{document}

\thispagestyle{empty}

\newcommand{\email}[1]{\vbox{\center\tt#1}\vspace{5mm}} 
\renewcommand{\thefootnote}{\fnsymbol{footnote}}
\setcounter{page}{1}
\setcounter{footnote}{0}
\setcounter{figure}{0}

\begin{titlepage}

\begin{center}

\vskip 2.3 cm 

\vskip 5mm

{\Large \bf
On the Spectrum of Pure Higher Spin Gravity
}

\vskip 0.5cm

\vskip 15mm

\centerline{Luis F. Alday$^{a}$, Jin-Beom Bae$^{a}$,  Nathan Benjamin$^{b}$, Carmen Jorge-Diaz$^{a}$}
\bigskip
\centerline{\it $^{a}$Mathematical Institute, University of Oxford, Woodstock Road, Oxford, OX2 6GG, UK.} 
\smallskip
\centerline{\it $^{b}$Princeton Center for Theoretical Science, Princeton University, Princeton, NJ 08544, USA.}

\vskip 1 cm

 \email{alday@maths.ox.ac.uk, bae@maths.ox.ac.uk \\ nathanb@princeton.edu, carmen.jorgediaz@maths.ox.ac.uk}

\end{center}

\vskip 2 cm

\begin{abstract}
\noindent  
We study the spectrum of pure massless higher spin theories in $AdS_3$. The light spectrum is given by a tower of massless particles of spin $s=2,\cdots,N$ and their multi-particles states. Their contribution to the torus partition function organises into the vacuum  character of the ${\cal W}_N$ algebra. Modular invariance puts constraints on the heavy spectrum of the theory, and in particular leads to negative norm states, which would be inconsistent with unitarity. This negativity can be cured by including additional light states, below the black hole threshold but whose mass grows with the central charge. We show that these states can be interpreted as conical defects with deficit angle $2\pi(1-1/M)$. Unitarity allows the inclusion of such defects into the path integral provided $M \geq N$.
\end{abstract}

\end{titlepage}


\setcounter{page}{1}
\renewcommand{\thefootnote}{\arabic{footnote}}
\setcounter{footnote}{0}

 \def\nref#1{{(\ref{#1})}}
 
 \tableofcontents
 
\section{Introduction}

Interacting theories of massless higher spin fields on $AdS_3$ have remarkable properties. Consistent theories where the graviton couples to a finite tower of higher spin particles appear to exist \cite{Blencowe:1988gj}, unlike what happens in higher dimensions.  As with the graviton, these higher spin fields do not have propagating degrees of freedom and these theories are `locally trivial'. The interesting dynamics is confined to the boundary and is described by a two dimensional conformal field theory. As shown in the seminal work by Brown and Henneaux \cite{Brown:1986nw} the asymptotic symmetries of standard gravity on $AdS_3$ are described by two copies of the Virasoro algebra with a central extension already at the classical level and whose central charge is given by
\begin{equation}
c_g=\frac{3 \ell}{2 G_N} 
\end{equation}
where $\ell$ is the radius of $AdS_3$ and $G_N$ the 3d Newton constant. In the case of higher spin gravity and for massless particles of spin $s=2,\cdots,N$ the asymptotic symmetries are described by two copies of the ${\cal W}_N$ algebra, with the same central charge \cite{Henneaux:2010xg,Campoleoni:2010zq}. It is believed that this algebra survives at the quantum level, see \cite{Gaberdiel:2010ar}. Although these theories possess a rich spectrum of classical solutions, see for instance \cite{Gutperle:2011kf}, it is natural to ask whether they provide consistent theories of quantum gravity, or they suffer from pathologies at the quantum level. Indeed, as discussed in \cite{Perlmutter:2016pkf}, it is hard to imagine how a finite tower of massless higher spins could arise from limits of string theory. This question can also be rephrased in terms of the dual CFT, and we can ask whether consistent unitary, irrational 2d CFTs with extended symmetry ${\cal W}_N$ exist. 

In this paper we study the spectrum of massless higher spin theories on thermal $AdS_3$. The partition function is highly constrained by consistency conditions, including unitarity, modular invariance, and ${\cal W}_N$ symmetry. We will follow two complementary approaches. In the first part of the paper we study the thermal partition function of a unitary conformal field theory with central charge $c$ and symmetry ${\cal W}_N$, in the irrational regime where $c>N-1$. In the simplest set-up, corresponding to `pure' higher spin gravity, the light spectrum of the theory arranges into the vacuum character of ${\cal W}_N$ and the question is what does modular invariance say about the heavy spectrum, above the BTZ threshold. We follow two procedures to construct spectral densities consistent with a given light spectrum, modular invariance, and ${\cal W}_N$ symmetry: the Poincare construction \cite{Maloney:2007ud,Keller:2014xba}  and the Rademacher construction \cite{Alday:2019vdr}. The resulting densities display some unphysical features. In particular:

a) The spectrum of energies at fixed spin is continuous, rather than discrete.\footnote{The Poincare construction in the Virasoro case leads to a degeneracy of $-6$ at the BTZ threshold. The Rademacher construction, and both constructions for $N>2$, are smooth at threshold.} 

b) For sufficiently low energy the resulting spectrum contains negative norm states, and hence is inconsistent with unitarity. 

An important comment is in order. In both constructions, Poincare and Rademacher, there are ambiguities in the heavy spectrum, consistent with both modular invariance and  ${\cal W}_N$ symmetry. An example of such an ambiguity is given by the Poincare construction starting with a heavy `seed'. These ambiguities allow for small (non-exponential) changes to the spectrum. This makes the problem a) not very serious, since it could be that such ambiguities render the spectrum discrete. On the other hand, one can check that the number of states with negative norm is exponentially large for large values of the central charge, even at finite spin. Hence problem b) appears to be more serious. This negativity was first observed in \cite{Benjamin:2019stq} for pure gravity, and in this paper we show that it persists for pure HS gravity. We argue that a `minimal' modification to the light spectrum that cures this problem is to add a tower of scalar operators with dimensions
\begin{equation}
\label{conical}
\Delta = \frac{c-N+1}{12}\left(1-\frac{1}{M^2} \right)+ \cdots,~~~M=N,N+1,\cdots.
\end{equation}
While unitarity imposes non-trivial constraints on the representations that can appear for $N >2$, see \cite{Afkhami-Jeddi:2017idc}, it turns out these operators are consistent with those bounds, so that in principle this is allowed by unitarity. 

In the second part of the paper we interpret these states in terms of HS gravity on $AdS_3$. For $N=2$ it has been shown that these states have the interpretation of a conical defect geometry, with deficit angle $2\pi(1-\frac{1}{M})$, see \cite{Benjamin:2020mfz}, and it has been argued that they must be included in the Euclidean path integral of 3d gravity. In this paper we show that this interpretation extends to general ${\cal W}_N$, by studying the one-loop partition function of HS gravity on a quotient of thermal $AdS_3$ by $\mathbb{Z}_M$. Quite remarkably we show that only conical defects with $M \geq N$ are consistent with unitarity, in agreement with the range in (\ref{conical}). For conical defects this bound is stronger than the unitarity bounds found in \cite{Afkhami-Jeddi:2017idc}.  As a byproduct we also compute the one-loop correction to the relation between the 2d central charge $c$, and the `gravitational' central charge $c_g$ that arises from the Brown-Henneaux procedure for HS gravity on $AdS_3$. We find
\begin{equation}
c= c_g +2N^3-N-1+ o(1)
\end{equation}
where $o(1)$ vanishes for large central charge. For $N=2$ we recover $c = c_g+13$, which has also been obtained in \cite{Cotler:2018zff,Cotler:2020ugk}, from both CFT and bulk points of view. 

This paper is organised as follows. In Section \ref{sec:spectra} we compute the Poincare and Rademacher sums of the $\mathcal{W}_N$ vacuum character. In Section \ref{sec:negativity}, we show these spectra have a density of states that is not positive-definite, and we describe a natural way to remove these negativities. In Section \ref{sec:conical}, we show these additional states can be naturally viewed as conical defects in the higher spin gravity theory, and provide some checks at one loop. Finally we conclude and discuss some possible future directions in Section \ref{sec:conclude}. 

\section{Modular densities in HS gravity}
\label{sec:spectra}

Consider a unitary irrational 2d CFT with $\mathcal{W}_N$ symmetry and central charge $c>N-1$. Its states are labelled by the conformal weights $(h,\bar h)$ and its torus partition function is given by
\begin{equation}
Z(q,\bar q) = \sum_{h,\bar h} q^{h-c/24} \bar{q}^{\bar h-c/24}
\end{equation}
where $q=e^{2\pi i \tau},\bar q=e^{-2\pi i \bar \tau}$ with $\tau$ the modulus of the torus. This partition function satisfies a list of constraints. First, the CFT possesses $\mathcal{W}_N$ symmetry. Hence its partition function on a torus can be decomposed into $\mathcal{W}_N$ characters. $\mathcal{W}_2$ corresponds to the Virasoro algebra, where the conserved current is the spin two stress-tensor. For $\mathcal{W}_N$ we have a tower of conserved currents, up to spin $N$. The generic characters for $\mathcal{W}_N$ algebras are not known. However, they are known in the case where the chemical potential for the higher spin currents is turned off. For a generic $W(d_1,\cdots, d_r)$ algebra the characters take the form
\begin{equation}
\chi_{vac}(q) = \frac{q^{-\hat c}}{\eta(q)^r} \prod_{i=1}^r \prod_{j=1}^{d_i-1}(1-q^j),~~~\chi_h(q) = \frac{q^{h-\hat c}}{\eta(q)^r}
\end{equation}
where $\hat c = \frac{c-r}{24}$. For simplicity we can assume all $d_i$ are different and in this notation $\mathcal{W}_N = \mathcal{W}(2,3,\cdots,N)$. Although our methods apply equally well to other cases, our main focus will be $\mathcal{W}_N$. In this case
\begin{equation}
\chi^{(N)}_{vac}(q) = \frac{q^{-\hat c}}{\eta(q)^{N-1}} \prod_{i=1}^{N-1} (1-q^i)^{N-i},~~~\chi^{(N)}_h(q) = \frac{q^{h-\hat c}}{\eta(q)^{N-1}}
\label{eq:vacdefinition}
\end{equation}
where $\hat c = \frac{c-(N-1)}{24}$. Our first assumption will be that the partition function can be decomposed into such characters
\begin{equation}
Z(q,\bar q) = \chi^{(N)}_{vac}(q) \bar \chi^{(N)}_{vac}(\bar q) + \sum_{h,\bar h} d_{h,\bar h} \chi^{(N)}_h(q)\bar \chi^{(N)}_{\bar h}(\bar q).
\end{equation}
We have assumed there is no other conserved currents (other than those of the $\mathcal{W}_N$ algebra). The integer numbers $d_{h,\bar h}$ count the multiplicities of $\mathcal{W}_N$ primaries (but we are only turning on the conformal dimensions $(h,\bar h)$ and no other chemical potential). The second assumption is that the theory is unitary. For a unitary theory the degeneracies $d_{h,\bar h}$ are non-negative. Furthermore, the representations that appear in the character decomposition must be unitary. This is non-trivial for ${\cal W}_N$ for $N>2$, see \cite{Afkhami-Jeddi:2017idc}, and it implies a gap between the vacuum and the next operator
\begin{equation}
h \geq h_{crit} = \frac{c-(N-1)}{24}\left( 1 - \frac{6 \floor*{\frac{N}{2}}}{N(N^2-1)} \right),
\end{equation}
and the same for $\bar h$. As shown in Appendix \ref{sec:higherspin}, the requirement of unitarity makes $\mathcal{W}_N$ symmetry stronger than $\mathcal{W}_2$, as expected. Our third assumption is that the partition function satisfies modular invariance. Modular invariance is generated by $\gamma \in PSL(2,\mathbb{Z})$ acting on the complex moduli of the torus as follows
\begin{equation}
\tau \to \gamma \tau = \frac{a \tau+b}{s \tau-r},~~~\bar \tau \to \gamma \bar \tau = \frac{a \bar \tau+b}{s \bar \tau-r},
\end{equation}
where $a,b,s,r$ are integers, $s$ is taken to be positive, and $a r+b s=-1$. Note that this condition implies $r,s$ are coprime: $(r,s)=1$. Since this transformation results in an equivalent torus, the partition function should be invariant
\begin{equation}
\label{modular}
Z(q,\bar q) = Z(\gamma q,\gamma \bar q),
\end{equation}
where it will be convenient to choose a parametrisation where 
\begin{eqnarray}
&q = e^{-\beta+ \frac{2\pi i r}{s}},~~~~~\gamma q = e^{-\frac{4\pi^2}{s^2 \beta} + \frac{2\pi i a}{s}}, \\ 
&\bar q = e^{-\bar \beta- \frac{2\pi i r}{s}},~~~~~\gamma \bar q = e^{-\frac{4\pi^2}{s^2 \bar \beta} - \frac{2\pi i a}{s}}.
\end{eqnarray}
Our last assumption is some spectrum of light operators, where by light we denote operators below the BTZ threshold at $h=\bar h=\hat c$. With these assumptions, we would like to construct the resulting partition function, or equivalently, the density $\rho(h,\bar h)$ of $\mathcal{W}_N$ primaries. To this end it is convenient to define a new partition function, also modular invariant, given by
\begin{eqnarray}
Z^p(q,\bar q)= \left( y^{1/2} \eta(q) \eta(\bar q) \right)^{N-1} Z(q,\bar q),
\end{eqnarray}
where $y=\text{Im}(\tau)$. This has the decomposition
\begin{eqnarray}
\label{Zprimaries}
Z^p(q,\bar q)= y^{\frac{N-1}{2}}\left( vac+\sum_{h,\bar h} q^{h-\hat c} \bar q^{\bar h-\hat c} \right) =  y^{\frac{N-1}{2}}\left( vac+\int dh d\bar h \rho(h,\bar h) q^{h-\hat c} \bar q^{\bar h-\hat c} \right),
\end{eqnarray}
where the sum runs over $\mathcal{W}_N$ primaries, the density $\rho(h,\bar h)$ is defined such that
\begin{eqnarray}
\int_{h_0}^{h_0+\Delta h}\int_{\bar h_0}^{\bar h_0+\Delta \bar h} dh d\bar h \rho(h,\bar h) 
\end{eqnarray}
is the number of primary states in the range $(h,\bar h) \in [h_0,h_0+\Delta h] \times [\bar h_0,\bar h_0+\Delta \bar h]$, and $vac$ is defined as
\be
vac = q^{-\hat c} \bar{q}^{-\hat c} \(\prod_{i=1}^{N-1}(1-q^i)^{N-i}\)\(\prod_{i=1}^{N-1}(1-\bar{q}^i)^{N-i}\).
\ee
For a discrete spectrum the density is a sum over delta functions. In terms of $Z^p(q,\bar q)$ modular invariance reads 
\begin{equation}
\label{Zpmodular}
vac+\int dh d\bar h \rho(h,\bar h) q^{h-\hat c} \bar q^{\bar h-\hat c} = \left(\frac{\gamma y}{y} \right)^{\frac{N-1}{2}} \left( \gamma\,vac+\int dh d\bar h \rho(h,\bar h) (\gamma q)^{h-\hat c} (\gamma \bar q)^{\bar h-\hat c}  \right).
\end{equation}

For a general modular transformation we have
\begin{equation}
\tau = x+i y = \frac{\beta}{2\pi} i + \frac{r}{s},~~~~\bar \tau = x-i y = -\frac{\bar \beta}{2\pi} i + \frac{r}{s}
\end{equation}
so that $y= \frac{\beta+\bar \beta}{4\pi}$, while $\gamma y = \frac{\pi}{s^2}\left(\frac{1}{\beta} + \frac{1}{\bar \beta} \right)$, which leads to 
\begin{equation}
\frac{\gamma y}{y} = \frac{1}{s^2} \frac{4\pi^2}{\beta \bar \beta} = \frac{1}{s \tau -r}\frac{1}{s \bar \tau -r}.
\end{equation}

We can follow two methods to construct such densities, which we briefly discuss. One is the Poincare construction. The other is the Rademacher construction. 

\subsection{Poincare construction}
The idea of the Poincare construction is very simple \cite{Maloney:2007ud,Keller:2014xba}. We start with a light `seed' of conformal weights $(h_0,\bar h_0)$ and then sum over all its $PSL(2,\mathbb{Z})$ images: 
\begin{equation}
Z_{\text{Poincare}}(\tau,\bar \tau) = \sum_{\gamma \,\in \, \mathbb{Z} \setminus PSL(2,\mathbb{Z})} \chi^{(N)}_{h_0}(\gamma \tau)\bar \chi^{(N)}_{\bar h_0}(\gamma \bar \tau).
\end{equation}

This sum is in principle divergent, but as show in \cite{Maloney:2007ud,Keller:2014xba} zeta-function regularisation can be used to define a partition function. Our aim here is to obtain explicit results for the $\mathcal{W}_N$ case. Since we are ultimately interested in the density of states, we find it convenient to used the method of kernels discussed in \cite{Benjamin:2020mfz}. For a given seed, the idea is to compute the contribution to the density of states coming from one of the $PSL(2,\mathbb{Z})$ images of the seed, and then sum over all images. It is convenient to work at the level of  $Z^p(q,\bar q)$ defined in (\ref{Zprimaries}) and define the modular kernel $\mathbb{K}_{h', h}^{(\mathtt{w}, \gamma)}$ as the solution to the density for the following problem:
\begin{equation}
\int_0^\infty dh' \mathbb{K}_{h', h}^{(\mathtt{w}, \gamma)} e^{2\pi i \tau h'} = (-i(s \tau - r))^{-\mathtt{w}} e^{2\pi i (\gamma \tau) h}.
\end{equation}

In words, the modular kernel expresses a transformed character (appropriately rescaled, as in (\ref{Zprimaries})) as a linear combination of untransformed characters. Because of the factors of $y$ present in (\ref{Zprimaries}), we need such expression for general weights $\mathtt{w}$. We will be interested in the case $\mathtt{w}=\frac{N-1}{2}$. The solution is given by:
\begin{align}
 \mathbb{K}^{(\mathtt{w},\gamma)}_{h'h} &= \epsilon(\mathtt{w},\gamma)\({2\pi\over s}\)\(\frac{h'}{h}\)^{\frac{\mathtt{w}-1}2} e^{{2\pi i\over s}(ah-rh')} I_{\mathtt{w}-1}\(\frac{4\pi\sqrt{-h'h}}s\)  
 \label{eq:kernelw}
\end{align}
where $\epsilon(\mathtt{w},\gamma)$ is a $h, h'$-independent phase (see (A.6) of \cite{Benjamin:2020mfz}).  We will now consider a seed with arbitrary weights $(h_0,\bar h_0)$ and compute the resulting density for weights $(h,\bar h)$. As in \cite{Keller:2014xba} it will be convenient to parametrise the seed and densities in terms of energies and spins, defined by
\begin{eqnarray}
&E = h_0 + \bar h_0 -2 \hat c,~~~J= h_0 - \bar h_0,\\
&e = h + \bar h -2 \hat c,~~~j= h - \bar h,
\end{eqnarray}
where recall $\hat c = \frac{c-N+1}{24}$. The seed will then produce a term in the partition function proportional to 
\be
q^{\frac{E+J}2} \bar{q}^{\frac{E-J}2}.
\label{eq:DefineEJ}
\ee
The Poincare sum of (\ref{eq:DefineEJ}) is then given by 
\begin{align}
\rho_{E,J}(e, j) &= \frac12 \sum_{\gamma \in \mathbb Z \backslash SL(2,\mathbb Z)} \mathbb{K}^{(\mathtt{w}, \gamma)}_{\frac{e+j}2, \frac{E+J}2}\mathbb{K}^{*(\mathtt{w}, \gamma)}_{\frac{e-j}2, \frac{E-J}2} \nn\\
&= \frac 12\sum_{s=1}^{\infty} \sum_{r \in (\mathbb Z/s \mathbb Z)^*} \sum_{n=-\infty}^{\infty} e^{{2\pi i\over s}(aJ-rj)} e^{2\pi i n j} \left[\ldots\right]\nn\\
&= \sum_{s=1}^{\infty} \frac{K(j, J; s)}{s^{2\mathtt w}} \rho_{E/s^2, J/s^2}^{s=1}(e,j).
\label{eq:PoincareEJ}
\end{align}
Some comments are in order in (\ref{eq:PoincareEJ}). In the first line, the factor of $\frac12$ is due to the Jacobian of switching from $(h, \bar h)$ to $(e,j)$. In the second line, we expanded the sum in $\mathbb Z\backslash SL(2,\mathbb Z)$, where we have taken $r'=r-ns$. The $\left[\ldots\right]$ refer to the remaining terms in the kernel (\ref{eq:kernelw}), i.e. $\(\frac{2\pi}s\)\(\frac{h'}{h}\)^{\frac{\mathtt{w}-1}2}I_{\mathtt{w}-1}\(\frac{4\pi\sqrt{-h'h}}s\)$ which we suppressed for brevity. The sum over $n$ simply projects onto integer spin $j$. Finally in the last line, $K(j, J; s)$ is a Kloosterman sum, defined as
\be
K(j, J; s) =  \sum_{r \in (\mathbb Z/s \mathbb Z)^*}  e^{{2\pi i\over s}(aJ-rj)}, ~~~~~-ar \equiv 1 ~(\text{mod} ~s)
\ee
and $\rho_{E/s^2, J/s^2}^{s=1}(e,j)$ is defined as
\begin{align}
\rho_{E,J}^{s=1}(e,j) &= 2\pi^2(-E+J)^{\frac{1-\mathtt{w}}2} (-E-J)^{\frac{1-\mathtt{w}}2} (e-j)^{\frac{\mathtt{w}-1}2} (e+j)^{\frac{\mathtt{w}-1}2} \nn\\
&\times I_{\mathtt{w}-1}\(2\pi\sqrt{-E+J}\sqrt{e-j}\)I_{\mathtt{w}-1}\(2\pi\sqrt{-E-J}\sqrt{e+j}\).
\label{eq:rhoEJ}
\end{align}
where as explained above $\mathtt{w}=\frac{N-1}{2}$. All in all, the Poincare density for a seed with quantum numbers $(E,J)$ for a theory with $\mathcal{W}_N$ symmetry is given by
\begin{equation}
\rho^{(N)}_{E,J}(e,j) = \sum_{s=1,2,3,\cdots} \frac{K(j,J,s)}{s^{N-1}} \rho_{E/s^2,J/s^2}^{(N),s=1}(e,j)
\end{equation}
with
\begin{align}
\label{rhopoincare}
\begin{split}
\rho_{E,J}^{(N),s=1}(e,j) &= 2\pi^2(-E+J)^{\frac{3-N}{4}}(-E-J)^{\frac{3-N}{4}}(e-j)^{\frac{N-3}{4}} (e+j)^{\frac{N-3}{4}} \\
                      &\quad \times I_{\frac{N-3}{2}} \left( 2\pi \sqrt{-E+J} \sqrt{e-j}\right) I_{\frac{N-3}{2}} \left( 2\pi \sqrt{-E-J} \sqrt{e+j}\right).
\end{split}
\end{align}
As a side remark, note that the convergence properties of the sum over images improves for large $N$. For $N>3$ the results are absolutely convergent. The case $N=3$ is discussed in Appendix \ref{sec:eisenstein}. 

\subsection{Rademacher construction}
We can also construct a density of states consistent with a seed $(E,J)$ and modular invariance by following the Rademacher procedure introduced in \cite{Alday:2019vdr}. The discussion for $\mathcal{W}_N$ very much mimics the discussion for $\mathcal{W}_2$, and we refer the reader to \cite{Alday:2019vdr} for the details. The idea is to look at the modular equation (\ref{Zpmodular}) in $x,y$ variables (where $\tau \equiv x+ i y, ~\bar\tau \equiv x - i y$):
\begin{equation}
vac + \sum_{j,e} \rho(e,j) e^{2\pi i j x} e^{-2\pi y e} = \left(\frac{1}{(s x-r)^2+s^2 y^2 }\right)^{\frac{N-1}{2}}\sum_{j',e'} e^{\frac{2\pi i a}{s} j'} e^{\frac{2\pi i(r-sx) j'}{s(r-sx)^2+s^3y^2}} e^{-\frac{2\pi y e'}{(r-sx)^2+s^2 y^2}}, 
\end{equation}
and assume $x,y$ are independent complex variables. We can then study this equation in the complex $(x,y)$ plane. These modular relations together with the existence of light operators imply a tower of essential singularities at points
\begin{equation}
x = \pm i y + \frac{r}{s}.
\end{equation}
We assume the existence of at least one such light operator, with quantum numbers $(E,J)$, which produces such a essential singularity
\begin{equation}
\label{modularEJ}
vac + \sum_{j,e} \rho(e,j) e^{2\pi i j x} e^{-2\pi y e} \sim \left(\frac{1}{(s x-r)^2+s^2 y^2 }\right)^{\frac{N-1}{2}}e^{\frac{2\pi i a}{s} J} e^{\frac{2\pi i(r-sx) J}{s(r-sx)^2+s^3y^2}} e^{-\frac{2\pi y E}{(r-sx)^2+s^2 y^2}}, 
\end{equation}
In order to proceed we rewrite the l.h.s. as 
\begin{equation}
\sum_{j,e} \rho(e,j) e^{2\pi i j x} e^{-2\pi y e}  = \sum_{j=-\infty}^\infty a_j(y) e^{2\pi i j x}
\end{equation}
where we have taken into account the spin is discrete and $a_j(y)$ encodes the contribution to the partition function from operators with a given spin. Following exactly the procedure in \cite{Alday:2019vdr} the relation (\ref{modularEJ}) leads to an expression for $a_j(y)$, which for $s=1$ is given by
\begin{align}
\label{aj expression}
a_j^{s=1}(y) = -i \int_{\frac{1}{2\pi y}-i \infty}^{\frac{1}{2\pi y }+ i\infty} dw \frac{(2\pi)^{N-2} w^{N-3}}{(4\pi w y -1)^{\frac{N-1}{2}}} e^{j(\frac{1}{w}-2\pi y)} e^{\frac{2\pi^2 w}{4\pi w y-1}\kappa_-} e^{2\pi^2 w \kappa_+},
\end{align}
where $\kappa_{\pm} = -(E \pm J)$ and $w= -\frac{1}{2\pi i(x+i y)}$. For $N=2$ this reduces to the integral encountered in \cite{Alday:2019vdr}, and for general $N$ can be computed in exactly the same way: expand in powers of $j$ and $\kappa_+$ and use the identity
\begin{align}
\frac{1}{2\pi i} \int_{\frac{1}{2\pi y}-i \infty}^{\frac{1}{2\pi y}+i \infty} dw \frac{w^{\alpha}}{(4\pi w y -1)^\beta} e^{2\pi^2 \kappa_+ w} = \frac{\kappa_+^{-\alpha+\beta-1} {_1\tilde{F}_1}\left(\beta;\beta-\alpha;\frac{\kappa_+ \pi}{2y}\right)}{2^{\alpha+\beta+1} \pi^{2\alpha-\beta+2} y^\beta},
\end{align}
term by term, where ${_1\tilde{F}_1}$ is regularized hypergeometric function. Once we obtain $a_j^{s=1}(y)$ as an expansion, we can find the resulting density $\rho(e,j)$ by an inverse Laplace transform. This procedure can be repeated for any value of $s$ and the final result again takes the form
\begin{equation}
\rho^{(N)}_{E,J}(e,j) = \sum_{s=1,2,3,\cdots} \frac{K(j,J,s)}{s^{N-1}} \rho_{E/s^2,J/s^2}^{(N),s=1}(e,j)
\end{equation}
where now 
\begin{align}
\label{rhorademacher}
\begin{split}
\rho_{E,J}^{(N),s=1}(e,j) &= 2\pi^2(-E+J)^{\frac{3-N}{4}}(-E-J)^{\frac{3-N}{4}}(e-j)^{\frac{N-3}{4}} (e+j)^{\frac{N-3}{4}} \\
                      &\quad \times I_{\frac{N-3}{2}} \left( 2\pi \sqrt{-E+J} \sqrt{e-j}\right) I_{-\frac{N-3}{2}} \left( 2\pi \sqrt{-E-J} \sqrt{e+j}\right).
\end{split}
\end{align}
The result is almost identical to that of the Poincare construction, except the inverse of the second Bessel function has the opposite sign. For odd $N$ the results actually agree, since for integer $n$ we have $I_{-n}(z) = I_{n}(z)$.  For even $N$ and in the regime where the density grows exponentially, the two results differ by an exponentially suppressed contribution since
\begin{equation}
I_\nu(z) \sim \frac{1}{\sqrt{2\pi z}} \left( e^z\left(1+ \frac{1-4\nu^2}{8z}+\cdots \right) - i e^{-i \pi \nu} e^{-z}\left(1 - \frac{1-4\nu^2}{8z}+\cdots \right) \right).
\label{eq:besselstuff}
\end{equation}
Both constructions, of course, have the Cardy behaviour as can be seen in the leading behaviour of (\ref{eq:besselstuff}) at large $z$.

\section{Negative norm states and how to cure them}
\label{sec:negativity}

We now will discuss if the spectra computed in Section \ref{sec:spectra} are consistent with the basic axioms of CFTs. The first comment we make is that much like the MWK spectra \cite{Maloney:2007ud, Keller:2014xba}, the spectra we compute here are continuous, rather than discrete sums of delta functions. This is indicative of them not being consistent with being the spectra of a single compact, unitary $\mathcal{W}_N$ CFT. However, it is possible they may be interpreted as an ensemble average of CFTs\footnote{See e.g. \cite{Afkhami-Jeddi:2020ezh, Maloney:2020nni} for recent discussions of a simpler setting of an ensemble average of CFTs with a holographic dual.}. In this section we will show that the spectra have a more serious problem, which is they are not positive definite. In particular, at certain spins and energies, they have an arbitrarily large negative number of states. This is very similar to the Virasoro case of the MWK partition function, as pointed out in \cite{Benjamin:2019stq}.

Let us first write the full density of states for either the Poincare or Rademacher sum of a $\mathcal{W}_N$ vacuum primary operator at central charge $c$. The $\mathcal{W}_N$ vacuum character is given by
\be
\chi_0(\tau) = \frac{q^{-\frac{c-N+1}{24}}}{\eta(\tau)^{N-1}} \prod_{i=1}^{N-1} (1-q^i)^{N-i}.
\ee
If we define
\be
\prod_{i=1}^{N-1}(1-q^i)^{N-i} = \sum_{i=0}^{\frac{N^3-N}6} c_N(i) q^i,
\label{eq:cdefN}
\ee
then the exact density of states is given by
\begin{align}
\rho^{(N)}(e, j) &= \sum_{i=0}^{\frac{N^3-N}6}\sum_{\bar i=0}^{\frac{N^3-N}6} c_N(i) c_N(\bar i) \rho_{-\frac{c-N+1}{12} + i + \bar i, i-\bar i}^{(N)}(e,j)
\label{eq:finalanswerPoincare}
\end{align}
where 
\be
\rho_{E, J}^{(N)}(e, j) = \sum_{s=1}^{\infty} \frac{K(j, J; s)}{s^{N-1}} \rho_{E/s^2, J/s^2}^{(N), s=1}(e,j) 
\label{eq:kloostermanstuff}
\ee
and $\rho_{E, J}^{(N), s=1}(e,j)$ is defined by either (\ref{rhopoincare}) or (\ref{rhorademacher}), for the Poincare and Rademacher sum respectively.

In the negativity we will discover, we need to take the limit as the energy approaches the spin, i.e. $e\rightarrow j$.\footnote{For convenience, in this section, we will always take $j\geq0$ without loss of generality. The density of states for negative $j$ is the same as the density of states for its absolute value.} This limit is taken before the limit of large central charge. The density of states in this limit drastically simplifies. In particular, for both the Poincare and Rademacher constructions, we have the scaling behavior
\be
\rho_{E, J}^{(N), s=1}(e,j) \sim (e-j)^{\frac{N-3}2}
\label{eq:n32}
\ee
as $e\rightarrow j$. However in (\ref{eq:finalanswerPoincare}), (\ref{eq:kloostermanstuff}), due to the properties of the function $c_N(i)$ and the Kloosterman sums, the $s=1, 2, \ldots N-1$ terms vanish more severely than (\ref{eq:n32}) in the limit $e\rightarrow j$. Therefore in that limit, the dominant contribution comes from the $s=N$ term in (\ref{eq:kloostermanstuff}). However, for $s>1$, the expression is no longer manifestly positive. We may ask how worrisome this negativity is. In other words, since we have a continuous density, we need to count the number of states with negative norm. This is simply given by the integral of the density over the range where it is negative. We obtain
\begin{equation}
\text{negative states} \sim \int_{j}^{j+e^{-\frac{8\pi}{N^2} \sqrt{\frac{c-N+1}{24} j}}}  (e-j)^{\frac{N-3}{2}} e^{\frac{4\pi}{N} \sqrt{\frac{c-N+1}{24} j}} de \sim e^{\frac{4\pi}{N^2} \sqrt{\frac{c-N+1}{24} j}}
\end{equation}
in particular, for either large central charge or large spin, the number of operators with negative norm is exponentially large. 

One natural way to cure such negativities is to include the Poincare or Rademacher sum of a sufficiently low-twist operator, where we define the \emph{twist} of an operator as min$(h, \bar h)$.\footnote{In \cite{Maxfield:2019hdt, Maxfield:2020ale}, it was argued that such a negativity can be also cured in a modular invariant way without introducing low-twist operators. It would be interesting to either prove or disprove that introducing a low-twist operator is the \emph{only} way to avoid such negativities.} In order to remove the negativity, we actually consider two separate regimes:
\begin{enumerate}
\item $j \gg c \gg 1$
\item $c \gg 1,~~~~~ j \ll c.$
\end{enumerate}
Region 1 will lead us to add an operator with $\text{min}(h, \bar h) \leq \frac{c-N+1}{24}\(1-\frac1{N^2}\)$. Region 2 is more subtle. It will lead us to add a certain number of states scaling as 
\be
\text{min}(h, \bar h) \leq \frac{c}{24}\(1-\frac1{N^2}\) + \mathcal{O}(1).
\label{eq:Scaling}
\ee
Moreover, if we saturate (\ref{eq:Scaling}), there is a minimum number of states we must add. In particular we will need to add a minimum of $\frac{N\phi(N)}{\text{gcd}(\phi(N), N-1)}$ states, where $\phi(N)$ is the Euler totient function. Finally, if we add the exact minimum number of states scaling as (\ref{eq:Scaling}), then the one-loop piece of (\ref{eq:Scaling}) is fully-determined as well. In particular, the \emph{average} of all states added must obey 
\begin{align}
\text{min}(h, \bar h) &\leq \frac{c-N+1}{24}\(1-\frac1{N^2}\) + \frac{N^2-1}{12N} = \frac{c+N+1}{24}\(1-\frac1{N^2}\).
\label{eq:ShiftC}
\end{align}
We illustrate this for the case of both $\mathcal{W}_2$ (Virasoro) and $\mathcal{W}_3$, and then discuss the general $\mathcal{W}_N$ case in the following sections.

\subsection{Example: $\mathcal{W}_2$ (Virasoro)}

The positivity of the density of states for the $SL(2,\mathbb Z)$ sum of the Virasoro vacuum character was analyzed in \cite{Benjamin:2019stq} for large spin, and in \cite{Alday:2019vdr} for finite spin; we restate the results here, focusing not on the large spin regime, but the regime where the spin is much less than the central charge. The vacuum character of the Virasoro algebra for $c>1$ is
\be
\chi_0(\tau) = \frac{e^{-2\pi i \tau\(\frac{c-1}{24}\)}}{\eta(\tau)}(1-q).
\ee

In our notation (\ref{eq:cdefN}), we would then have
\be
c_2(0)=1, ~ c_2(1)=-1.
\ee 

In the limit of $e\rightarrow j$ and large $c$ (where we first take $e\rightarrow j$), the density of states (\ref{eq:finalanswerPoincare}, \ref{eq:kloostermanstuff}) is dominated by the smallest $s$ such that the answer is nonzero. In particular there must exist some spin mod $s$ and some $k=0, 1$ where
\be
c_2(k) \sum_{i=0}^{1} \frac{K(j, k-i; s)c_2(i)}{s}
\label{eq:thiscanbenegativeW2}
\ee 
is nonzero. In the limit of large $c$, (\ref{eq:thiscanbenegativeW2}) will then multiply $\exp\(2\pi\sqrt{\(\frac{c-1}{24}-k\)j}\)$. This is the same behaviour we see for a non-degenerate Virasoro primary of twist $\frac{c-1}{24}\(1-\frac1{s^2}\)+\frac{k}{s^2}$. For $s=1$, (\ref{eq:thiscanbenegativeW2}) always vanishes. We summarize the values of (\ref{eq:thiscanbenegativeW2}) for $s=2$ in the Table \ref{table:w2}.

\begin{table}[ht]
\begin{center}
\begin{tabular}{ |c|c|c|c|p{3cm}|} 
 \hline
$k$ & Multiplying what? & $j$ even & $j$ odd & ~~Twist to add\\ 
\hline
0& $\exp\(2\pi\sqrt{\(\frac{c-1}{24}\)j}\)$ & $+1$ & $-1$ & $\frac{c-1}{24}\(1-\frac1{2^2}\)$  \\ 
1& $\exp\(2\pi\sqrt{\(\frac{c-1}{24} - 1\)j}\)$ & $+1$ & $-1$ & $\frac{c-1}{24}\(1-\frac1{2^2}\) + \frac14$  \\ 
\hline
\multicolumn{2}{|c|} {\bf Average:} & (positive) & {\footnotesize $\frac{-1(0)-1(\frac14)}2=-\frac18$} & $\frac{c-1}{24}\(1-\frac1{2^2}\) + \frac18$ (two states) \\
\hline
\end{tabular}
\end{center}
\caption{In this table, we look at the negativity in the Virasoro vacuum character's modular sum. The entries in the third and fourth columns evaluate (\ref{eq:thiscanbenegativeW2}) at $s=2$ (the smallest value of $s$ such that (\ref{eq:thiscanbenegativeW2}) is non-vanishing) for the values of $k$ and $j$ indicated. The second column indicates the magnitude of this term to the partition function at large $j$ or large $c$, and the last column indicates the twist necessary to add in order to obtain a comparable density of states from a non-vacuum state. We see there is negativity for odd spins. At finite $j$ and large $c$, for both of these, we can cure the negativity by introducing two states with $\text{min}(h, \bar h) \leq \frac{c-1}{24}\(1-\frac1{2^2}\)+\frac18$.}
\label{table:w2}
 \end{table} 

We then see from Table \ref{table:w2} that the density of states goes roughly as
\be
\rho_j(e) \sim - \frac{e^{2\pi \sqrt{\(\frac{c-1}{24}\)j}}}{\sqrt{e-j}} - \frac{e^{2\pi\sqrt{\(\frac{c-1}{24}-1\)j}}}{\sqrt{e-j}}
\ee
for odd $j$ when we take the limit $e\rightarrow j$. If we add the $SL(2,\mathbb Z)$ sum of two primary operators with dimensions scaling as $\text{min}(h, \bar h) = \frac{c-1}{32} + \frac{t_i}4$, then the density of states in this limit looks like
\be
\rho_j(e) \sim - \frac{e^{2\pi \sqrt{\(\frac{c-1}{24}\)j}}}{\sqrt{e-j}} - \frac{e^{2\pi\sqrt{\(\frac{c-1}{24}-1\)j}}}{\sqrt{e-j}} + \sum_{i=1}^2 \frac{e^{2\pi\sqrt{\(\frac{c-1}{24} - t_i \)j}}}{\sqrt{e-j}}.
\label{eq:posgoal}
\ee 
We now ask: in (\ref{eq:posgoal}), for what values of $t_i$ is the expression positive? If we take the limit of large spin, i.e. $j \gg c$, this implies that there is at least one operator with $t_i \leq 0$. Instead let us take the limit of $j \ll c$. In that limit (\ref{eq:posgoal}) becomes
\be
\rho_j(e) \sim \frac{e^{2\pi\sqrt{\(\frac{c-1}{24}\)j}}}{\sqrt{e-j}}\(-\pi \sqrt{\frac{24j}{c-1}}(t_1 + t_2 -1) + \mathcal{O}(c^{-1})\)
\ee
which is non-negative at this order in the $1/c$ expansion if the average of the $t_i$'s we add is less than or equal to $\frac12$. This means that the states we added obey
\be
\text{min}(h, \bar h) \leq \frac{c-1}{24}\(1-\frac1{2^2}\) + \frac18.
\label{eq:zxdv}
\ee
The value in (\ref{eq:zxdv}) (including the one-loop shift) has a very natural gravitational interpretation as a conical defect geometry explored in \cite{Benjamin:2020mfz}. We will review this in Section \ref{sec:conical}, and generalize to the case of $\mathcal{W}_N$. Note that if we added more than two states we would not get any nontrivial condition on the twist of the added operators in the regime where $j \ll c$. 

\subsection{Example: $\mathcal{W}_3$}

Let us now redo this computation for the case of $\mathcal{W}_3$. The vacuum character of the $\mathcal{W}_3$ algebra for $c>2$ is
\be
\chi_0(\tau) = \frac{e^{-2\pi i \tau\(\frac{c-2}{24}\)}}{\eta(\tau)^2}(1-2q+2q^3-q^4).
\ee
Again, in our notation (\ref{eq:cdefN}), we would then have
\be
c_3(0)=1, ~ c_3(1)=-2, ~c_3(2)=0, ~c_3(3)=2, ~c_3(4)=-1.
\ee
In the limit of $e\rightarrow j$ and large $c$, the density of states (\ref{eq:finalanswerPoincare}, \ref{eq:kloostermanstuff}) is dominated by the smallest $s$ such that the answer is nonzero. In particular there must exist some spin mod $s$ and some $k=0, 1, \ldots, 4$ where
\be
c_3(k) \sum_{i=0}^{4} \frac{K(j, k-i; s)c_3(i)}{s^2}
\label{eq:thiscanbenegative}
\ee
is nonzero. In the limit of large $c$, (\ref{eq:thiscanbenegative}) will then multiply $\exp\(\frac{4\pi\sqrt{\(\frac{c-2}{24}-k\)j}}{s}\)$. This is the same behavior we see for a non-degenerate $\mathcal{W}_3$ primary of twist $\frac{c-2}{24}\(1-\frac1{s^2}\)+\frac{k}{s^2}$. For $s=1, 2$ (\ref{eq:thiscanbenegative}) always vanishes. We summarize the values of (\ref{eq:thiscanbenegative}) for $s=3$ in the Table \ref{table:w3}.

\begin{table}[ht]
\begin{center}
\begin{tabular}{ |c|c|c|c|c|p{3cm}|} 
 \hline
$k$ & Multiplying what? & $j \equiv 0 ~(\text{mod}~3)$ & $j \equiv 1 ~(\text{mod}~3)$ & $j \equiv 2 ~(\text{mod}~3)$ &~~Twist to add\\ 
\hline
0& $\exp\(\frac{4\pi\sqrt{\(\frac{c-2}{24}\)j}}{3}\)$ & $+1$ & $-1$ & $0$ & $\frac{c-2}{24}\(1-\frac1{3^2}\)$  \\ 
1& $\exp\(\frac{4\pi\sqrt{\(\frac{c-2}{24}-1\)j}}{3}\)$ & $+2$ & $0$ & $-2$ & $\frac{c-2}{24}\(1-\frac1{3^2}\) + \frac19$  \\ 
2& $\exp\(\frac{4\pi\sqrt{\(\frac{c-2}{24}-2\)j}}{3}\)$ & $0$ & $0$ & $0$ & $\frac{c-2}{24}\(1-\frac1{3^2}\) + \frac29$  \\  
3& $\exp\(\frac{4\pi\sqrt{\(\frac{c-2}{24}-3\)j}}{3}\)$ & $+2$ & $-2$ & $0$ & $\frac{c-2}{24}\(1-\frac1{3^2}\) + \frac39$  \\ 
4& $\exp\(\frac{4\pi\sqrt{\(\frac{c-2}{24}-4\)j}}{3}\)$ & $+1$ & $0$ & $-1$ & $\frac{c-2}{24}\(1-\frac1{3^2}\) + \frac49$  \\ 
\hline
\multicolumn{2}{|c|} {\bf Average:} & (positive) &  {\footnotesize $\frac{-1(0)-2(\frac39)}3=-\frac29$} & {\footnotesize $\frac{-2(\frac19)-1(\frac49)}3=-\frac29$} & $\frac{c-2}{24}\(1-\frac1{3^2}\) + \frac29$ (three states) \\
\hline
\end{tabular}
\end{center}
\caption{In this table, we look at the negativity in the $\mathcal{W}_3$ vacuum character's modular sum. The entries in the third, fourth, and fifth columns evaluate (\ref{eq:thiscanbenegative}) at $s=3$ (the smallest value of $s$ such that (\ref{eq:thiscanbenegative}) is non-vanishing) for the values of $k$ and $j$ indicated. The second column indicates the magnitude of this term to the partition function at large $j$ or large $c$, and the last column indicates the twist necessary to add in order to obtain a comparable density of states from a non-vacuum state. We see there is negativity for spins $j \equiv 1, 2~ (\text{mod}~3$). At finite $j$ and large $c$, for both of these, we can cure the negativity by introducing three states with $\text{min}(h, \bar h) \leq \frac{c-2}{24}\(1-\frac1{3^2}\)+\frac29$.}
\label{table:w3}
 \end{table}
 
In the limit where $j\gg c$, the most important term is the largest exponential in the spin, which occurs for $j\equiv 1~(\text{mod}~3)$. Therefore we need to add a state with $\text{min}(h, \bar h) \leq \frac{c-2}{24}\(1-\frac{1}{3^2}\)$. However, this is not sufficient. For spins $j$ that are much smaller than the central charge, we have not sufficiently cured all the negativity in Table \ref{table:w3}. For spins $j \equiv 1, 2~(\text{mod}~3)$, we need three states with an average shift of $\frac29$. In other words, we need three states with an average twist of
\be
\text{min}(h, \bar h) \leq \frac{c-2}{24}\(1-\frac{1}{3^2}\) + \frac29.
\label{eq:goalw3}
\ee
To see this, let us take $j\equiv 1~(\text{mod}~3)$ as an example. Suppose we add the Poincare sum of three states with twists
\be
\text{min}\(h, \bar h\) = \frac{c-2}{24}\(1-\frac1{3^2}\)+\frac{t_i}9 + \mathcal{O}(c^{-1})
\ee
for $i=1, 2, 3$, and where each of the $t_i$'s are $\mathcal{O}(1)$ numbers. Then the total density of states at spin $j\equiv 1~(\text{mod}~3)$ for $j\ll c$ is 
\begin{align}
\rho_j &\sim - e^{\frac{4\pi}{3}\sqrt{\(\frac{c-2}{24}\)j}} -2 e^{\frac{4\pi}{3}\sqrt{\(\frac{c-2}{24} - 3\)j}} + \sum_{i=1}^3 e^{\frac{4\pi}{3}\sqrt{\(\frac{c-2}{24} - t_i\)j}} \nn\\
&\sim e^{\frac{4\pi}3\sqrt{\(\frac{c-2}{24}\)j}}\((4\pi - \frac{2\pi(t_1+t_2+t_3)}{3})\sqrt{\frac{j}{\frac{c-2}{24}}} + \mathcal{O}\(c^{-1}\)\)
\label{eq:asdfgh}
\end{align}
which is positive if the average of the $t_i$'s is at most $2$, reproducing (\ref{eq:goalw3}). Note that if we added fewer than three states, then at sufficiently large $c$, the density of states would become negative. We find similar results by considering $j\equiv 2~(\text{mod}~3)$.

\subsection{$\mathcal{W}_N$}

More generally, for $\mathcal{W}_N$, we can repeat the same calculation. The expression
\be
c_N(k) \sum_{i=0}^{\frac{N^3-N}{6}} \frac{K(j, k-i; s)c_N(i)}{s^{N-1}}
\label{eq:thiscanbenegativeWN}
\ee 
is nonzero for $s\geq N$. In particular, if we minimize over $j$, we find that
\begin{align}
\text{min}_j\sum_{i=0}^{\frac{N^3-N}6} \sum_{k=0}^{\frac{N^3-N}6} \frac{c_N(i) c_N(k) K\(j, k-i; N\)}{N^{N-1}} = -\frac{N\phi(N)}{\text{gcd}(\phi(N), N-1)}
\label{eq:AllEqN}
\end{align}
 where $\phi(N)$ is the Euler totient function. We can now compute the average shift necessary to cure the negativity after introducing $\frac{N\phi(N)}{\text{gcd}(\phi(N), N-1)}$ primary operators, each with twist scaling as (to leading order in $c$) $\text{min}(h, \bar h) \sim \frac{c}{24}\(1-\frac1{N^2}\)$. We find that
\begin{align}
\(-\frac{N\phi(N)}{\text{gcd}(\phi(N),N-1)}\)^{-1}\sum_{i=0}^{\frac{N^3-N}6} \sum_{k=0}^{\frac{N^3-N}6} \(\frac{c_N(i) c_N(k) K\(j, k-i; N\)}{N^{N-1}}\)\times\(\frac k{N^2}\) = \frac{N^2-1}{12N}
\label{eq:AllEq}
\end{align}
where in (\ref{eq:AllEq}), $j$ is given by any $j$ such that (\ref{eq:AllEqN}) is minimized.\footnote{We have checked (\ref{eq:AllEqN}, \ref{eq:AllEq}) up to $N=16$. It would be interesting to prove both analytically.}

%

Thus, we will find that we need states with an average twist of 
\be
\text{min}(h, \bar h) \leq \frac{c-N+1}{24}\(1-\frac{1}{N^2}\) + \frac{N^2-1}{12N}.
\ee 

\section{Conical defects in HS gravity}
\label{sec:conical}
In previous sections we have seen that the partition function of `pure' HS gravity on $AdS_3$ possesses states of negative norm. A scenario to cure this negativity is to consider extra operators below the BTZ threshold, whose twist behaves as
\begin{equation}
\tau_M  = \frac{c-(N-1)}{24} \left( 1-\frac{1}{M^2} \right) + \cdots,~~~~M \geq N.
\end{equation}
In the case of pure gravity on $AdS_3$, corresponding to a CFT with Virasoro symmetry, it has been argued that such states have the interpretation of conical singularities, which should be added to the path integral \cite{Benjamin:2020mfz}.  In this section, we consider conical singularities in HS gravity, and compute their contribution to the Euclidean partition function. As an aside, we pause to note that the addition of such conical defects has appeared in top-down constructions of $AdS_3$ gravity. It was shown in \cite{Lunin:2002bj} that when one considers strings on $AdS_3 \times S^3 \times T^4$, conical defect geometries show up in the partition function as twisted sector ground states of the symmetric product orbifold. This identification of conical defect geometries was recently made especially sharp in the tensionless limit of string theory in \cite{Eberhardt:2020bgq}. 

We are interested in computing the one loop Euclidean partition function for quadratic fluctuations around thermal $AdS_3$
\begin{equation}
Z(q) = \int [{\cal D} \varphi] e^{-S[\varphi]},
\end{equation}
where $q=e^{2\pi i \tau}$, with $\tau$ the complex structure of the torus at the boundary of thermal $AdS_3$. For now we are only summing over smooth geometries\footnote{More precisely HS gravity on $AdS_3$ is described in terms $SL(N, \mathbb R) \times SL(N, \mathbb R)$ connections, and smoothness of the metric, which is not a gauge invariant concept, is replaced by triviality of the holonomies around the thermal circle.}. $S[\varphi]$ is the action for a tower of  massless higher spin fields $\varphi_{(s)}$, with $s=2,\cdots,N$. The quadratic part of the action is simply the sum of individual actions  $S[\varphi_{(s)}]$. $S[\varphi_{(2)}]$ is the familiar gravity action on $AdS_3$, while $S[\varphi_{(s)}]$ is the Fronsdal action for a massless spin-s field in $AdS_3$. It then follows that the partition function is the product of spin-s partition functions
\begin{equation}
Z (q)= \prod_{s=2}^N Z^{(s)}(q).
\end{equation}
The one-loop partition function $Z^{(s)}(q)$ for spin two was computed in \cite{Giombi:2008vd}, while for generic spin $s$ was computed in \cite{Gaberdiel:2010ar}. It takes the form
\begin{equation}
\label{partitions}
Z^{(s)}(q) = \left( q \bar q\right)^{-\frac{1}{24}\epsilon(s)} \prod_{n=s}^\infty \frac{1}{(1-q^n)(1-\bar q^n)},~~~q= e^{2\pi i \tau},
\end{equation}
with $\epsilon(s)$ a one-loop shift of the central charge, which is $\epsilon(2)=13$ for spin two, see \cite{Giombi:2008vd,Cotler:2018zff}, and will be given below for general spin. Plugging this into the expression for the partition function we find
\begin{equation}
Z (q)=  \left( q \bar q\right)^{-\frac{c_g}{24}+\frac{shift}{24}}  \prod_{s=2}^N \prod_{n=s}^\infty \frac{1}{(1-q^n)(1-\bar q^n)} \sim \chi_{vac}^{(N)}(q) \chi_{vac}^{(N)}(\bar q),
\end{equation}
where we have included the tree level contribution $\left( q \bar q\right)^{-\frac{c_g}{24}}$ together with the resulting shift of the central charge. Here $c_g$ is the gravitation central charge that arises from the Brown-Henneaux procedure. Up to a one-loop shift, this exactly agrees with the vacuum character of ${\cal W}_N \times \bar {\cal W}_N$ algebra. This is of course expected, since this is the algebra of asymptotic isometries. We will now proceed as follows. First we will show how to reproduce (\ref{partitions}) using the heat kernel method. With this method we will then generalise the result to non-smooth geometries, corresponding to the presence of conical singularities. 

\subsection{The heat kernel}
Suppose we want to compute the following partition function to one-loop
\begin{equation}
Z = \int {\cal D}{\varphi}e^{- \frac{1}{\hbar^2} \int_{\cal M} d^3 x \sqrt{g} \varphi \Delta \varphi}
\end{equation}
for a free scalar field $\varphi$ and where $\Delta$ is the Laplacian operator on ${\cal M}$. This is simply a Gaussian integral and the result can be given in terms of the determinant of $\Delta$. For compact ${\cal M}$ the Hilbert space admits a discrete basis in terms of the eigenfunctions of $\Delta$:
\begin{equation}
\Delta \varphi_n(x) = \lambda_n \varphi_n(x).
\end{equation}
It is often the case that such a basis is complete, and can also be chosen to be orthonormal, so that
\begin{equation}
\sum_n \varphi_n(x) \varphi_n(y) = \delta(x,y),~~~~\int_{\cal M} d^3x \sqrt{g} \varphi_m(x) \varphi_n(x) = \delta_{mn} ,
\end{equation}
where $\delta(x,y)$ is the appropriate delta-function on ${\cal M}$ and for simplicity we have assumed the basis is real. Then the determinant is given by
\begin{equation}
S^{(1)} = -\frac{1}{2} \log \det \Delta = -\frac{1}{2} \sum_n \log \lambda_n.
\end{equation}
The heat kernel method provides a very powerful way to compute this.\footnote{See \cite{David:2009xg} for a thorough review, and \cite{Giombi:2008vd} for one of the original applications to the present context.} Introduce the heat kernel
\begin{equation}
K(t;x,y) = \langle y| e^{-t \Delta} |x \rangle=\sum_n \varphi_n(x) \varphi_n(y) e^{-\lambda_n t} .
\end{equation}
Then using the completeness relation it follows
\begin{equation}
S^{(1)} = \frac{1}{2} \lim_{\alpha \to 0} \partial_\alpha \int_0^\infty dt \frac{t^{\alpha-1}}{\Gamma(\alpha)}  \int_{\cal M} d^3 x \sqrt{g} K(t;x,x).
\end{equation}
The advantage of writing the determinant in terms of the heat kernel is that $K(t;x,y)$ can also be defined as the unique solution to the following problem
\begin{equation}
\left(\partial_t +\Delta_x \right) K(t;x,y) =0,~~~K(0,x,y) = \delta(x,y),
\end{equation}
and it is usually simpler to solve this rather than the original problem. Before we move on to the case of interest, a few comments are in order. First, the method applies to fields with spin. In that case the heat kernel has to be supplemented with spin indices $K(t;x,y) \to K_{a,b}(t;x,y)$. Second, for the case of interest ${\cal M}$ will be non-compact. In this case we have a continuous spectrum and the determinant will contain a divergence proportional to the volume of ${\cal M}$. Third, once the heat kernel is computed for a given manifold ${\cal M}$ it is in principle straightforward to compute the heat kernel for quotient manifolds ${\cal M}/\Gamma$. We will see examples of this below. 

\bigskip
\noindent
{\bf Thermal $AdS_3$}
\bigskip

\noindent
We are interested in the heat kernel for a spin $s$ field in Euclidean $AdS_3$, denoted by  $H_3^{+}$, and more specifically its thermal version, given by the quotient $H_3^{+}/\mathbb{Z}$. The metric of $H_3^{+}$ is given by
\begin{equation}
ds^2 = \frac{dz d\bar z+dy^2}{y^2},~~~y>0.
\end{equation}
On these coordinates the generator of $\mathbb{Z}$ acts as 
\begin{equation}
\gamma \cdot (y,z,\bar z) = (|q|^{-1} y, q^{-1} z, \bar q^{-1} \bar z),
\end{equation}
where $q=e^{2\pi i \tau}$. This leaves the metric invariant but is equivalent to a global identification, so that the boundary is a torus of modulus $\tau=\tau_1+i \tau_2$. The corresponding heat kernel for generic spin $s$ has been worked out in \cite{Gaberdiel:2010ar}, but we find it convenient to apply the geometrical approach of \cite{David:2009xg}. Following their notation we introduce the integrated kernel for thermal $AdS_3$ given by
\begin{equation}
K^{(s)}(t,\tau,\bar \tau) = \sum_{m \in \mathbb{Z}} \sum_a \int_{H_3^{+}/\mathbb{Z}} d^3 x \sqrt{g} K_{aa}^s(t;x,\gamma^m \cdot x).
\end{equation}
The sum over $m$ corresponds to the sum over images that defines the quotient. The indices $a,b$ in $K_{a,b}(t;x,y)$ are labels for the $2s+1$ dimensional spin-s representation. Exploiting the fact that $H_3^{+} \cong \frac{SL(2,\mathbb{C})}{SU(2)}$ in \cite{David:2009xg} an expression for $K^{(s)}(t,\tau,\bar \tau)$ is given, in terms of $SL(2,\mathbb{C})$ characters $\chi_{j_1,j_2}(\alpha)$ for the diagonal group element $diag(\alpha,\alpha^{-1}) \in SL(2,\mathbb{C})$:
\begin{align}
\label{character formula}
\chi_{j_1, j_2}(\alpha) = \frac{\alpha^{2j_1+1} \bar{\alpha}^{2j_2+1} + \alpha^{-2j_1-1} \bar{\alpha}^{-2j_2-1}}{(\alpha-\alpha^{-1})(\bar{\alpha}-\bar{\alpha}^{-1})}.
\end{align}
In terms of these the expression takes the form
\begin{equation}
K^{(s)}(t,\tau,\bar \tau) =K^{(s)}_0(t) + \tau_2 \sum_{m \in \mathbb{Z}/\{0\}} \int_0^\infty d\lambda \left( \chi_{j_1,j_2}(e^{m \pi i \tau})+ \chi_{j_2,j_1}(e^{m \pi i \tau}) \right) e^{-t(\lambda^2+s+1)},
\end{equation}
where $2j_1=s-1+i \lambda,~2j_2=-s-1+i \lambda$ and the contribution from $m=0$ is given by \cite{David:2009xg}
\begin{equation}
\label{zero mode heat kernel}
K^{(s)}_0(t)  = \frac{\text{vol}\left(H_3^{+}/\mathbb{Z} \right)}{(4\pi t)^{3/2}}(2-\delta_{s,0})(1+2 s^2 t) e^{-t(s+1)}.
\end{equation}
Performing the integral over $\lambda$ we obtain
\begin{equation}
\label{nonzero mode heat kernel}
K^{(s)}(t,\tau,\bar \tau) =K^{(s)}_0(t)+ \tau_2 \sum_{m=1} \frac{\cos(2\pi s m \tau_1 )}{\sin(m \pi \tau)\sin(m \pi \bar \tau)} \frac{\sqrt{\pi}}{\sqrt{t}} e^{-\frac{\pi^2 m^2 \tau_2^2}{t}-t(s+1)}
\end{equation}
As shown in \cite{Gaberdiel:2010ar} the one-loop partition function for the spin-s contribution is given by
\begin{align}
\log Z^{(s)}(\tau) &= -\frac{1}{2}\log \det \left( \Delta_s - s(s-3) \right) + \frac{1}{2}\log \det \left( \Delta_{s-1} - s(s-1) \right) \cr
 &=  \frac{1}{2} \lim_{\alpha \to 0} \partial_\alpha \int_0^\infty dt \frac{t^{\alpha-1}}{\Gamma(\alpha)}  \left(K^{(s)}(t,\tau,\bar \tau) e^{-s(s-3)t}- K^{(s-1)}(t,\tau,\bar \tau) e^{-s(s-1)t} \right) \nonumber \\
 &= \frac{6s-6s^2-1}{6\pi}\text{vol}\left(H_3^{+}/\mathbb{Z} \right)+\sum_{m=1}^\infty \frac{1}{m} \left( \frac{q^{m s}}{1-q^m} +  \frac{\bar q^{m s}}{1-\bar q^m} \right).
\end{align}
The volume $\text{vol}\left(H_3^{+}/\mathbb{Z} \right)$ is divergent, but following \cite{Cotler:2018zff} we can regularise it and obtain
\begin{equation}
\text{vol}\left(H_3^{+}/\mathbb{Z} \right) = - \tau_2 \pi^2
\end{equation}
Putting all together we obtain
\begin{equation}
\label{1-loop of boson}
Z^{(s)}_{1-loop}(\tau)  = \left( q \bar q\right)^{-\frac{1}{24}(6s^2-6s+1)} \prod_{n=s}^\infty \frac{1}{(1-q^n)(1-\bar q^n)},
\end{equation}
in perfect agreement with the result quoted at the beginning of this section . Considering now the total contribution to the one-loop partition function arising from spins $s=2,3,\cdots,N$ we find
\begin{equation}
Z (q)=  \left( q \bar q\right)^{-\frac{1}{24}(c_g-1-N+2N^3)}  \prod_{s=2}^N \prod_{n=s}^\infty \frac{1}{(1-q^n)(1-\bar q^n)}.
\end{equation}
Note that this reproduces precisely the Vacuum ${\cal W}_N$ character provided the central charge of the conformal field theory is corrected by the shift
\begin{equation}
\label{2dcharge}
c = c_g-1-N+2 N^3,
\end{equation}
where $c_g$ is the classical central charge resulting from the Brown-Henneaux procedure. 

As an aside, we pause to note that \eqref{2dcharge} will take a qualitatively different form if we redo the calculation with supersymmetry. For $\mathcal{N}=(1,1)$ supersymmetry, the relation between $c$ and $c_g$ is quadratic in $N$ instead of cubic; for $\mathcal{N}=(2,2)$ supersymmetry it is linear in $N$; and finally for more amounts of supersymmetry we have $c=c_g$. See Appendix \ref{sec:SUGRAshift} for details. 

\bigskip
\noindent
{\bf Conical defects}
\bigskip

\noindent
We are now ready to compute the one-loop partition function in the presence of a conical defect singularity.\footnote{The conical defects considered in this paper are different to the ones considered in \cite{Castro:2011iw}. Theirs correspond to smooth geometries,  which in the $SL(N)$ Chern-Simons formulation of higher spin gravity in $AdS_3$ correspond to trivial holonomies for the $SL(N,\mathbb R)$ connection. For the solutions considered here, the holonomy to the $M^{\text{th}}$ power is trivial, but the holonomy itself is not. Furthermore, the solutions considered here are always consistent with unitarity, unlike the solutions in \cite{Castro:2011iw}.} The conical defect background is given by a further quotient, $H_3^{+}/\Gamma$ with $\Gamma=\mathbb{Z} \times \mathbb{Z}_M$. We have now two generators $\gamma_1,\gamma_2$ that act as
\begin{equation}
\gamma_1 \cdot (y,z,\bar z) = (|q|^{-1/M} y, q^{-1/M} z, \bar q^{-1/M} \bar z),~~~\gamma_2 \cdot (y,z,\bar z) = (y, e^{-2\pi i/M} z, e^{2\pi i/M} \bar z).
\end{equation}
 Note that $\gamma_2^M=1$. The heat kernel can again be computed by the method of images and takes the form
\begin{equation}
K_M^{(s)}(t,\tau,\bar \tau) =K^{(s)}_{M,0}(t) + \frac{\tau_2}{M^2} \sum_{n=1}^M \sum_{m \in \mathbb{Z}/\{0\}} \int_0^\infty d\lambda \left( \chi_{j_1,j_2}(e^{\pi i \tau\frac{m}{M}}e^{\pi i\frac{n}{M}})+ \chi_{j_2,j_1}(e^{\pi i \tau\frac{m}{M}}e^{\pi i\frac{n}{M}}) \right) e^{-t(\lambda^2+s+1)}.
\end{equation}
with $j_1,j_2$ as before, and we have singled out the contribution from $m=0$. Let us first analyse this piece. This will now have two terms
\begin{eqnarray}
K^{(s)}_{M,0}(t)  &=& \frac{\text{vol}\left(\frac{H_3^{+}}{\mathbb{Z} \times \mathbb{Z}_M}\right)}{(4\pi t)^{3/2}}(2-\delta_{s,0})(1+2 s^2 t) e^{-t(s+1)} +\frac{\sqrt{\pi}\tau_2}{2\sqrt{t} M^2} \sum_{n=1}^{M-1} \frac{\cos \left(\frac{2\pi s n}{M}\right)}{\sin^2\left(\frac{n \pi}{M}\right)}e^{-t(s+1)}\\
&=& \left( \frac{\text{vol}\left(\frac{H_3^{+}}{\mathbb{Z} \times \mathbb{Z}_M}\right)}{(4\pi t)^{3/2}}(2-\delta_{s,0})(1+2 s^2 t)   + \frac{\sqrt{\pi}\tau_2}{6\sqrt{t} M^2}(6s^2+M^2-6 M s-1)\right) e^{-t(s+1)} \nonumber.
\end{eqnarray}
As before, the divergent volume of $\frac{H_3^{+}}{\mathbb{Z} \times \mathbb{Z}_M}$ can be regularised and we obtain
\begin{equation}
\text{vol}\left(\frac{H_3^{+}}{\mathbb{Z} \times \mathbb{Z}_M}\right) = -\frac{\pi^2 \tau_2}{M^2}.
\end{equation}
With this we obtain
\begin{eqnarray}
K^{(s)}_{M,0}(t)  =\frac{\sqrt{\pi }\tau_2  \left(2 t \left(M^2-6 M s+3 s^2-1\right)-3\right)}{12 M^2 t^{3/2}} e^{-(s+1) t}.
\end{eqnarray}
Going back to $K_M^{(s)}(t,\tau,\bar \tau)$ and performing the integrals over $\lambda$ we obtain
\begin{eqnarray}
K_M^{(s)}(t,\tau,\bar \tau) = K^{(s)}_{M,0}(t)  +\frac{\sqrt{\pi} \tau_2}{2 \sqrt{t} M^2} \sum_{n=1}^M \sum_{m \in \mathbb{Z}/\{0\}}  \frac{\cos\left(\frac{2\pi s}{M}(n+m \tau_1)\right)}{\sin \left( \frac{\pi}{M} (n+m \tau)\right)\sin \left( \frac{\pi}{M} (n+m \bar \tau)\right)} e^{-\frac{m^2\pi^2 \tau_2^2}{M^2 t}-(s+1)t}.
\end{eqnarray}
Plugging this into the expression for the one-loop determinant we obtain

\begin{eqnarray}
\log Z^{(s)}(\tau) &=& \frac{1}{2} \lim_{\alpha \to 0} \partial_\alpha \int_0^\infty dt \frac{t^{\alpha-1}}{\Gamma(\alpha)}  \left(K_M^{(s)}(t,\tau,\bar \tau) e^{-s(s-3)t}- K_M^{(s-1)}(t,\tau,\bar \tau) e^{-s(s-1)t} \right) \\
&=& \frac{\pi}{6}\tau_2 - \sum_{n=1}^M \sum_{m=1}^\infty \sum_{\ell=0}^\infty \frac{1}{m M} \cos \left( \frac{2\pi n(\ell+s)}{M} \right)\left( q^{\frac{m}{M}(\ell+s)} +\bar q^{\frac{m}{M}(\ell+s)} \right) 
\end{eqnarray}
The sum over $m$ can be done at once, and the sum over $n$ gives
\begin{eqnarray}
\sum_{n=1}^M \cos \left( \frac{2\pi n(\ell+s)}{M} \right) = M \delta_{0, \ell+s \Mod{M}}.
\end{eqnarray}
This implies $\ell+s = k M$, but note the smallest value of $k$ is $\floor*{\frac{s-1}{M}}+1$. Putting together all ingredients we obtain
\begin{equation}
Z^{(s)}_{1-loop}(\tau) = (q \bar q)^{-\frac{1}{24}} \prod_{k=\floor*{\frac{s-1}{M}}+1} \frac{1}{(1-q^k)(1-\bar q^k)}.
\end{equation}
Note that the dependence on $M$ and $s$ is only through the floor $\floor*{\frac{s-1}{M}}$. Let's consider now the total contribution from fields of spin $s=2,3,\cdots,N$ to the one-loop partition function. We will first assume $N \leq M$ so that all products in $Z^{(s)}_{1-loop}(\tau) $ start from $k=1$. In this case we obtain
\begin{equation}
Z^{(s),N}(\tau) = (q \bar q)^{-\frac{c_g}{24 M^2}-\frac{N-1}{24}} \prod_{s=2}^N \prod_{k=1}^\infty \frac{1}{|1-q^k|^2} =\frac{ (q \bar q)^{-\frac{c}{24 M^2}} }{|\eta(\tau)|^{2(N-1)}},
\end{equation}
where we have included the classical contribution, which in the presence of a conical singularity is $ (q \bar q)^{-\frac{c_g}{24 M^2}}$. This exactly agrees with a ${\cal W}_N$ non-degenerate character for a scalar operator. What is the weight of this operator in the CFT? Comparing the expressions for the non-degenerate characters and denoting $c$ the central charge of the CFT we find $h=\bar h$ with
\begin{equation}
h-\frac{c-(N-1)}{24} = -\frac{c_g}{24 M^2},
\end{equation}
and rewriting the BH central charge $c_g$ in terms of $c$, using (\ref{2dcharge}) we find
\begin{equation}
h= \frac{c-N+1}{24} \left(1-\frac{1}{M^2} \right) + \frac{N^3-N}{12 M^2}.
\end{equation}
Let us end with the following comment. Imagine we consider $N > M$. This will lead to terms of the form $(1-q)(1-\bar q)$ missing from the final expression for  $Z^{(s),N}(\tau)$. In order to write the result in terms of non-degenerate ${\cal W}_N$ characters, we need to divide and multiply by the corresponding factors, so that now the answer will be, for instance, of the form $(1-q)(1-\bar q)\frac{ (q \bar q)^{-\frac{c_g}{24 M^2}} }{|\eta(\tau)|^2}$. This however corresponds to two scalar operators with positive norm, together with two operators with $J=\pm 1$ and negative norm, and hence it seems to violate unitarity. We conclude that unitarity requires $N \leq M$.

\bigskip
\noindent
{\bf Comparison}
\bigskip

\noindent
The twist of the first allowed conical defect is given by
\begin{equation}
\tau= \frac{c-N+1}{24} \left(1-\frac{1}{N^2} \right) + \frac{N^2-1}{12 N}.
\end{equation}
Which agrees precisely with the average of operators needed to cure negativity in section \ref{sec:negativity}. We can now ask the following question. Imagine we add this contribution to the path integral, with the degeneracy derived in the CFT computation. To what extent does this cure the negativity? Let us consider ${\cal W}_2$ for simplicity and work in the regime where $c \gg j$. The density in the dangerous region is given by
\begin{equation}
\rho_j(e) \sim -e^{2\pi \sqrt{\left(\frac{c-1}{24}\right) j}}-e^{2\pi \sqrt{\left(\frac{c-1}{24}-1\right) j}} +e^{2\pi \sqrt{\left(\frac{c-1}{24}-\frac12 + \delta \tau_1 \right) j}} +e^{2\pi \sqrt{\left(\frac{c-1}{24}-\frac12 + \delta \tau_2 \right) j}} ,
\end{equation}
where $\delta \tau_1,\delta \tau_2$ are higher loop corrections to the twist of the conical defects, and in particular could lift the degeneracy. These corrections have two sources: higher order corrections to the relation between the 2d central charge and the gravitational central charge $c_g$; and quantum corrections to the conical defects themselves. We would like to make a few observations. First, if $\delta \tau_i \sim \frac1c$, as naively expected, then the addition of the two conical defects above does not quite cure the negativity problem. On the other hand, for $\delta \tau_i  > \frac1{c^{1/2}}$, then quantum effects become sufficiently important, and negativity could be cured. The situation is very similar for all HS theories. There are many reasons why the $1/c$ expansion could contain non-analytic terms (such as fractional powers or terms containing $\log c$). This could arise for instance, from a higher dimensional embedding necessary to render the theory UV complete or when regularising divergences. 
\section{Conclusions}

\label{sec:conclude}

In this paper we analyzed both the Poincare and Rademacher sums of the $\mathcal{W}_N$ vacuum characters for $c>N-1$. We found that both constructions, similar to the Virasoro case, result in a continuous spectrum of $\mathcal{W}_N$ primary operators that is not positive definite. One natural way to cure such negativity is the addition of the Poincare/Rademacher sum of operators with twist parametrically below $c/24$. We then interpreted these operators as $\mathbb Z_M$ orbifolds of the higher spin gravity theory in AdS$_3$. 

This proposal passes several nontrivial checks. The first is that the Poincare/Rademacher sum of the $\mathcal{W}_N$ character is negative, and from a CFT perspective this negativity is naturally cured with states that obey
\be
\text{min}(h, \bar h) \leq \frac{c}{24}\(1-\frac1{N^2}\) + \mathcal{O}(1),
\ee
which would correspond to $\mathbb Z_N$ orbifolds in the higher spin gravity theory. We indeed then find that these are precisely the first orbifolds that are consistent to include in the gravitational path integral.

The second consistency check is the one-loop piece of the story. If we demand positivity in the spins with $j \ll c$, then we are compelled to add states that obey in average
\be
\text{min}(h, \bar h) \leq \frac{c-N+1}{24}\(1-\frac1{N^2}\) + \frac{N^2-1}{12N},
\label{eq:blahblahpigpig}
\ee
which is precisely the one-loop shift that we compute for the $\mathbb Z_N$ orbifolds from the gravity computation via the heat kernel method. 

In order to compare the one-loop shift of the states added to the CFT in (\ref{eq:blahblahpigpig}) with the one-loop shift of the $\mathbb Z_N$ orbifolds from the higher spin gravity theory, it was necessary to relate the Brown-Henneaux central charge $c_g = \frac{3\ell_{\text{AdS}}}{2G_N}$ with the CFT central charge $c$:
\begin{equation}
c= c_g +2N^3-N-1+ o(1).
\end{equation} 
This generalizes the computation done for Virasoro to the case of $\mathcal{W}_N$.

There are a number of puzzles and interesting questions that deserve mention. The first is it is worth pausing to emphasize that it is still an open question as to whether or not any theory with $\mathcal{W}_N$ symmetry (and no enhanced chiral algebra) exists for $c>N-1$. Indeed in \cite{Afkhami-Jeddi:2017idc}, it was shown that unitarity of the Kac matrix places lower bounds that scale with $c$ on the twist of \emph{all} non-vacuum primary operators in such putative theories. Our bounds (\ref{eq:blahblahpigpig}) are consistent with the unitarity bounds of \cite{Afkhami-Jeddi:2017idc}, but it would be interesting to explore whether further constraints can rule out (or identify) these CFTs. A candidate for such constraints is positivity and discreteness/integrability of the spectrum. As discussed in \cite{Alday:2019vdr} both the Rademacher or Poincare construction, which only take modular invariance plus a given light spectrum into account, suffer from ambiguities. We expect much of these ambiguities to be fixed by the requirement of positivity and discreteness/integrability of the spectrum. See \cite{Kaidi:2020ecu} for a discussion along these lines. Perhaps in the context of higher spin theories one could show that these conditions, plus the extra HS symmetry, are either too strong, and in this way one could rule out these theories, or help fix the spectrum completely.   

Another point we would like to emphasize is other potential ways to resolve the negativity seen in the density of states. In recent work \cite{Maxfield:2019hdt, Maxfield:2020ale}, it was proposed that a spin-dependent shift in the BTZ threshold could be enough to cure the negativity in a modular invariant way. This was obtained by proposing that a new class of topologies should be included in the path integral, which do not correspond to classical solutions. It would be interesting to see if a similar analysis argues a way to fix the negativity in higher spin gravity; and also to explicitly construct positive modular invariant partition functions with states essentially only appearing at the BTZ threshold (both for Virasoro and for $\mathcal{W}_N$).

Finally, we have seen that the issue of negativity after adding the first tower of conical defects is sensitive to higher order loop corrections. The naive expectation is that both the relation between CFT and gravity central charges, as well as the dimension of conical defects, receive corrections of order $1/c$. However, as discussed in the previous section, this may not be the case. It would be very interesting to consider higher loop corrections to the path integral.  

\color{black}

\section*{Acknowledgements} 
We would like to thank Alejandra Castro, Scott Collier, Lorenz Eberhardt, and Alex Maloney for very helpful discussions. This project has received funding from the European Research Council (ERC) under the European Union's Horizon 2020 research and innovation programme (grant agreement No 787185). NB is supported in part by the Simons Foundation Grant No. 488653. CJD is supported by a Schreder Music Award.

\appendix

\section{Modular invariant partition functions}
\label{sec:eisenstein}
In this appendix we would like to show that the Poincare and Rademacher constructions both lead to a modular invariant partition function. In both constructions modular invariance is built in, but infinite sums are involved, which could in principle diverge. Here we would like to address this point. We will focus in the case of a scalar seed $J=0$, for which we have found a closed form expression. Taking the Poincare density (\ref{rhopoincare}) for $s=1$, expanding it in powers of $E$ and computing the contribution from a given spin $j$ to the partition function $Z^p$ we find 
\begin{equation}
\left. Z^p_{\text{Poincare}} \right|_{j}= \sum_{s=1} \frac{K(j,0,s)}{s^{N-1}} \sum_{k=0} \frac{2^{k+1} \sqrt{y} \left( -\frac{E}{s^2}\right)^{k} \pi ^{\frac{1}{2} (4 k+N-1)} j^{k+\frac{N}{2}-1} K_{k+\frac{N}{2}-1}(2 j \pi  y)}{\Gamma (k+1) \Gamma \left(k+\frac{N}{2}-\frac{1}{2}\right)}.
\end{equation}
We would like to write the full answer for the partition function in terms of known modular functions. The building blocks are the real analytic Eisenstein series 
\begin{equation}
E_m(\tau) = \frac{1}{2} \sum_{(p,q)=1} \frac{y^m}{|p \tau+q|^{2m}}.
\end{equation}
These have the following Fourier decomposition 
\begin{equation}
E_m(\tau) = y^m + \frac{\hat \zeta(2m-1)}{\hat \zeta(2m)} y^{1-m} + \frac{4}{\hat \zeta(2m)} \sum_{j=1} \cos(2\pi j x)j^{m-1/2} \sigma_{1-2m}(j) y^{1/2} K_{m-1/2}(2\pi j y),
\end{equation}
where $\sigma_{1-2m}(j)$ is the divisor function and 
\begin{equation}
\hat \zeta(z) = \pi^{-z/2} \Gamma \left( \frac{z}{2}\right) \zeta(z).
\end{equation}
The claim is that for a scalar seed the partition function $Z^p_{Poinc}(q,\bar q)$ can be written as a linear combination of Eisenstein series. To see this we write the divisor function $\sigma_{1-2m}(j)=j^{1-2m} \sigma_{2m-1}(j)$ as 
\begin{equation}
\sigma_{1-2m}(j) =\zeta(2m) \sum_{s=1}^\infty \frac{1}{s^{2m}} K(j,0;s).
\end{equation}
Comparing the contribution for a given spin $j$ we find 
\begin{equation}
Z^P_{\text{Poincare}}(\tau) = \sum_{k=0}^\infty \frac{(2 \pi) ^k (-E)^k}{\Gamma (k+1)}E_{\frac{N+2k-1}{2}}(\tau) .
\end{equation}
The partition functions arising from the Rademacher construction, for a scalar seed, can also be written as linear combinations of Eisenstein series. For $N$ odd the result agrees with that of the Poincare series. For $N$ even we obtain
\begin{equation}
Z^P_{\text{Rademacher}}(\tau) = \sum_{k=0}^\infty \frac{(2 \pi) ^{k-\frac{N}{2}+\frac{3}{2}} (-E)^{k-\frac{N}{2}+\frac{3}{2}}}{\Gamma \left(k-\frac{N}{2}+\frac{5}{2}\right)} E_{k+1}(\tau) .
\end{equation}
Let us now discuss the issue of convergence. We have used the density for $j \neq 0$ to express the partition function in terms of real Eisenstein series. For $N=2$ the issue of regularisation for the Poincare series was discussed in \cite{Maloney:2007ud}. For $N=3$ for both constructions, and for even $N$ in the Rademacher construction, we see the series contain the Eisenstein series $E_{1}(\tau)$, which is divergent. We can regularise it by using the first Kronecker limit formula which gives
\begin{equation}
E_{k}(\tau) = \frac{\pi}{k-1} + 2\pi (\gamma-\log 2- \log \left(\sqrt{y}|\eta(\tau)|^2 \right))  + {\cal O}(k-1).
\end{equation}
The divergent piece is a constant, $x-$independent, and hence it can be subtracted without spoiling modular invariance and without modifying the density for $j \neq 0$. On the other hand, the term that goes like $\log y$ cannot be subtracted. This will induce a density at spin zero found by computing the inverse Laplace transform of $y^{-\frac{N-1}{2}} \log y$. This leads to 
\begin{equation}
\rho(e) \sim e^{\frac{N-3}{2}} \left( \log e + c \right).
\end{equation}
As $e \to 0$ this diverges as a power for $N=2$, logarithmically for $N=3$ and converges for $N=4,6,\cdots$. \footnote{The density at spin zero is given by $\rho(e) \sim e^{\frac{N-3}{2}}$ for $N=5,7,\cdots$, therefore it converges as $e \to 0$.} 

\section{Higher spin symmetry and unitarity}
\label{sec:higherspin}
Since the characters for $\mathcal{W}_N$ algebra for any $N$ represent a complete set of holomorphic functions, one can wonder how strong the constraint that a given partition function admits a decomposition into $\mathcal{W}_N$ characters is. As we will see, when combined with unitarity, this can give very strong constraints. First, the representations that appear must be unitary. While for Virasoro algebra $\mathcal{W}_2$ all representations with $h \geq 0$ are unitary, this is not true in general. For general $\mathcal{W}_N$ the vacuum representation $h=0$ is unitary, and then there is a gap, and unitarity requires \cite{Afkhami-Jeddi:2017idc}
\begin{equation}
h \geq h_{crit} = \frac{c-(N-1)}{24}\left( 1 - \frac{6 \floor*{\frac{N}{2}}}{N(N^2-1)} \right).
\end{equation}
Second, the multiplicities must be positive. Consider for example a non-vacuum character for $\mathcal{W}_3$ and expand it into $\mathcal{W}_2$ characters. We obtain
\begin{equation}
\chi_h^{(3)}(q) = \chi_h^{(2)}(q)+\chi_{h+1}^{(2)}(q) +2\chi_{h+2}^{(2)}(q) +3\chi_{h+3}^{(2)}(q) + \cdots = \sum_{n=0} a(n) \chi_{h+n}^{(2)}(q)
\end{equation}
where $a(n)$ is positive and denotes the number of partitions of $n$. The same is true for the vacuum character for $\mathcal{W}_3$, which admits a decomposition into the vacuum plus non-vacuum characters of $\mathcal{W}_2$ with positive coefficient. Hence, if a partition function admits a decomposition into $\mathcal{W}_3$ characters of unitary representations with positive coefficients, it also admits a decomposition  into $\mathcal{W}_2$ characters of unitary representations with positive coefficients. The converse is not true. For instance, decomposing a $\mathcal{W}_2$ character into $\mathcal{W}_3$ we obtain 
\begin{equation}
\chi_h^{(2)}(q) = \chi_h^{(3)}(q)-\chi_{h+1}^{(3)}(q) -\chi_{h+2}^{(3)}(q) + \chi_{h+5}^{(3)}(q) + \chi_{h+7}^{(3)}(q) -\chi_{h+12}^{(3)}(q)+ \cdots 
\end{equation}
which contains negative coefficients. The same is true for the vacuum character.

\section{One-loop shift for HS supergravity}
\label{sec:SUGRAshift}

Our goal of this appendix is to show non-existence of the one-loop shift of the central charge when we impose a sufficient amount of supersymmetry. To see this, we apply the heat kernel method to compute the one-loop partition function of HS supergravity on thermal AdS$_3$. HS supergravity with $\mathcal{N}=(p,q)$ supersymmetry is known to be described by a Chern-Simons theory with an extended higher-spin superalgebra shs$^E$($p,q;\mathbb{C}$) that was constructed in \cite{Vasiliev:1986qx}. The theory also involves massive scalar particles as well as fermionic degrees of freedom coupled to the higher spin fields. Here we focus on the one-loop partition function of the massless particles. We first investigate the one-loop shift of the central charge for $\mathcal{N}=(1,1)$ and $\mathcal{N}=(2,2)$ HS supergravity, and then generalize the computation to the higher supersymmetric case.\footnote{We pause to note that we are computing the one-loop determinant formally, without worrying about whether or not such CFTs with these chiral algebras are even consistent. Indeed unitarity may place nontrivial bounds on such theories in the finite $N$, large $c$ limit (see e.g. \cite{Romans:1991wi,Banados:2015tft}.)}

\subsubsection*{$\mathcal{N}=(1,1)$ HS supergravity}

The one-loop partition function of the $\mathcal{N}=(1,1)$ HS supergravity takes the form
\begin{align}
\label{1-loop for SUGRA}
Z_{1-loop}^{(s)}(\tau,\bar{\tau}) = Z_{1-loop}^{(s),B}(\tau,\bar{\tau}) Z_{1-loop}^{(s-1),F}(\tau,\bar{\tau}) ,
\end{align}
where $Z_{1-loop}^{(s),B}(\tau,\bar{\tau})$ denotes the contribution of the massless bosonic particles of spin $s$ which was already computed in the main text in \eqref{1-loop of boson}. $Z_{1-loop}^{(s),F}(\tau,\bar{\tau})$ represents the one-loop partition function of the massless fermionic particles of spin $s+\frac{1}{2}$ and its explicit form is given by \cite{Creutzig:2011fe}
\begin{align}
\label{fermion one loop}
\log Z^{(s),F}_{1-loop}(\tau,\bar{\tau}) = \frac{1}{2} \log \text{det} \left(\frac{ \Delta_{s+\frac{1}{2}} - (s+\frac{1}{2})(s-\frac{5}{2})}{ \Delta_{s-\frac{1}{2}} - (s-\frac{1}{2})(s+\frac{1}{2})}\right)^{\text{TT}}.
\end{align}
With the help of the expressions of the heat kernel \eqref{zero mode heat kernel} and \eqref{nonzero mode heat kernel}, it is easy to see that the one-loop partition function for the spin $s+\frac{1}{2}$ fermion is given by
\begin{align}
\label{fermion}
Z^{(s),F}_{1-loop}(\tau,\bar{\tau}) = (q \bar{q})^{\frac{12s^2-1}{48}} \prod_{n=s}^{\infty} |1+q^{n+\frac{1}{2}}|^2.
\end{align}
We imposed a factor of $(-1)^m$ to the non-zero mode, due to the anti-periodicity of the fermionic fields along the thermal circle.

As a consistency check, let us apply above formulas to the $\mathcal{N}=(1,1)$ supergravity by simply substituting $s=2$ to \eqref{1-loop for SUGRA}. One can check that the one-loop partition function takes the form
\begin{align}
Z_{1-loop}^{s=2}(\tau, \bar{\tau}) = (q\bar{q})^{-\frac{c_g}{24}-\frac{5}{16}} \prod_{n=2}^{\infty} \frac{|1+q^{n-\frac{1}{2}}|^2}{|1-q^n|^2}.
\label{eq:oneLoopN1}
\end{align}
The above one-loop partition function (\ref{eq:oneLoopN1}) is simply the vacuum character of the $\mathcal{N}=1$ super-Virasoro algebra (with a renormalized central charge). It counts the thermal excitations obtained by acting with the super-Virasoro generators on the vacuum state. (\ref{eq:oneLoopN1}) therefore reproduces the one-loop partition function as  argued in \cite{Maloney:2007ud}.

To obtain the full one-loop partition function of the $\mathcal{N}=(1,1)$ HS supergravity, we consider the total contributions of spin $s=2,3,\cdots,N$. We find 
\begin{align}
\begin{split}
Z (\tau, \bar{\tau})  &=  \prod_{s=2}^N \left( Z^{(s),B}_{1-loop} (\tau, \bar{\tau}) Z^{(s-1),F}_{1-loop} (\tau, \bar{\tau}) \right) \\
       &=  \left( q \bar q\right)^{-\frac{1}{24}(c_g-\frac{3}{2} (1-N) + 3N^2)}  \prod_{s=2}^N \prod_{n=s}^\infty \frac{(1+q^{n-\frac{1}{2}})(1+\bar q^{n-\frac{1}{2}})}{(1-q^n)(1-\bar q^n)}
\end{split}
\end{align}
and this result implies that the one-loop shift of central charge is as follows:
\begin{equation}
c = c_g-\frac{3}{2} (1-N) + 3N^2
\label{eq:SUSYshiftc}
\end{equation}
The dependence on $N$ in the shift (\ref{eq:SUSYshiftc}) is quadratic, as opposed to the bosonic case \eqref{2dcharge}, where it was cubic.

\subsubsection*{$\mathcal{N}=(2,2)$ HS supergravity}

Let us now consider $\mathcal{N}=(2,2)$ HS supergravity. After considering the total contributions of the massless higher spin particles, we arrive at the expression \cite{Creutzig:2011fe}
\begin{align}
\begin{split}
Z_{1-loop}(\tau, \bar{\tau}) &= \prod_{s=2}^{N} Z^{(s),B}_{1-loop}(\tau, \bar{\tau}) Z^{(s-1),B}_{1-loop}(\tau, \bar{\tau})  \left(Z^{(s-1),F}_{1-loop}(\tau, \bar{\tau})\right)^2.
\end{split}
\end{align}
Immediately we find the one-loop partition function can be computed as
\begin{align}
\begin{split}
Z_{1-loop} &= (q \bar{q})^{-\frac{c_g}{24}-\frac{N-1}{8}} \prod_{s=2}^{N}  \left(\prod_{n=s}^{\infty} \frac{1}{|1-q^{n}|^2} \right) \left(\prod_{n=s-1}^{\infty} \frac{1}{|1-q^{n}|^2} \right)  \left(  \prod_{n=s-1}^{\infty} |1+q^{n+\frac{1}{2}}|^2 \right)^2,
\end{split}
\end{align}
and the one-loop shift of the central charge reads
\begin{align}
c = c_g + 3N-3.
\end{align}

\subsubsection*{HS supergravity with higher supersymmetry}

Finally, we consider the one-loop partition function for HS supergravity with higher supersymmetry. More precisely, we consider $\mathcal{N}=(p,p)$ supersymmetry for $p > 2$. In general, the partition function takes the form
\begin{align}
\label{partition for sugra}
\begin{split}
Z_{1-loop}(\tau, \bar{\tau}) &= \prod_{s=2}^{N} \left[ \prod_{i=0}^{\floor{\frac{p}{2}}} \left( Z^{(s-i),B}_{1-loop}(\tau, \bar{\tau}) \right)^{C(p,2i)} \prod_{j=0}^{\floor{\frac{p-1}{2}}} \left( Z^{(s-1-j),F}_{1-loop}(\tau, \bar{\tau}) \right)^{C(p,2j+1)} \right]
\end{split}
\end{align}
where $C(n, r)$ denotes the binomial coefficient. To proceed, we simply substitute \eqref{1-loop of boson} and \eqref{fermion} to \eqref{partition for sugra}. After some computation we find that the central charge reads
\begin{align}
c = c_g,
\end{align}
for any $p>2$. Therefore, we conclude that there is no one-loop shift in the central charge when we have enough supersymmetry.


\begin{thebibliography}{}

\bibitem{Blencowe:1988gj}
M.~P.~Blencowe,
``A Consistent Interacting Massless Higher Spin Field Theory in $D$ = (2+1),''
Class. Quant. Grav. \textbf{6} (1989), 443
doi:10.1088/0264-9381/6/4/005

\bibitem{Brown:1986nw}
J.~D.~Brown and M.~Henneaux,
``Central Charges in the Canonical Realization of Asymptotic Symmetries: An Example from Three-Dimensional Gravity,''
Commun. Math. Phys. \textbf{104} (1986), 207-226
doi:10.1007/BF01211590

\bibitem{Henneaux:2010xg}
M.~Henneaux and S.~J.~Rey,
``Nonlinear $W_{\infty}$ as Asymptotic Symmetry of Three-Dimensional Higher Spin Anti-de Sitter Gravity,''
JHEP \textbf{12} (2010), 007
doi:10.1007/JHEP12(2010)007
[arXiv:1008.4579 [hep-th]].

\bibitem{Campoleoni:2010zq}
A.~Campoleoni, S.~Fredenhagen, S.~Pfenninger and S.~Theisen,
``Asymptotic symmetries of three-dimensional gravity coupled to higher-spin fields,''
JHEP \textbf{11} (2010), 007
doi:10.1007/JHEP11(2010)007
[arXiv:1008.4744 [hep-th]].

\bibitem{Gaberdiel:2010ar}
M.~R.~Gaberdiel, R.~Gopakumar and A.~Saha,
``Quantum $W$-symmetry in $AdS_3$,''
JHEP \textbf{02} (2011), 004
doi:10.1007/JHEP02(2011)004
[arXiv:1009.6087 [hep-th]].


\bibitem{Gutperle:2011kf}
M.~Gutperle and P.~Kraus,
``Higher Spin Black Holes,''
JHEP \textbf{05} (2011), 022
doi:10.1007/JHEP05(2011)022
[arXiv:1103.4304 [hep-th]].


\bibitem{Perlmutter:2016pkf}
E.~Perlmutter,
``Bounding the Space of Holographic CFTs with Chaos,''
JHEP \textbf{10} (2016), 069
doi:10.1007/JHEP10(2016)069
[arXiv:1602.08272 [hep-th]].
    
\bibitem{Maloney:2007ud}
A.~Maloney and E.~Witten,
``Quantum Gravity Partition Functions in Three Dimensions,''
JHEP \textbf{02} (2010), 029
doi:10.1007/JHEP02(2010)029
[arXiv:0712.0155 [hep-th]].

\bibitem{Keller:2014xba}
C.~A.~Keller and A.~Maloney,
``Poincare Series, 3D Gravity and CFT Spectroscopy,''
JHEP \textbf{02} (2015), 080
doi:10.1007/JHEP02(2015)080
[arXiv:1407.6008 [hep-th]].

\bibitem{Alday:2019vdr}
L.~F.~Alday and J.~B.~Bae,
``Rademacher Expansions and the Spectrum of 2d CFT,''
JHEP \textbf{11}, 134 (2020)
doi:10.1007/JHEP11(2020)134
[arXiv:2001.00022 [hep-th]].


\bibitem{Benjamin:2019stq}
N.~Benjamin, H.~Ooguri, S.~H.~Shao and Y.~Wang,
``Light-cone modular bootstrap and pure gravity,''
Phys. Rev. D \textbf{100} (2019) no.6, 066029
doi:10.1103/PhysRevD.100.066029
[arXiv:1906.04184 [hep-th]].

\bibitem{Afkhami-Jeddi:2017idc}
N.~Afkhami-Jeddi, K.~Colville, T.~Hartman, A.~Maloney and E.~Perlmutter,
``Constraints on higher spin CFT$_{2}$,''
JHEP \textbf{05} (2018), 092
doi:10.1007/JHEP05(2018)092
[arXiv:1707.07717 [hep-th]].

\bibitem{Benjamin:2020mfz}
N.~Benjamin, S.~Collier and A.~Maloney,
``Pure Gravity and Conical Defects,''
JHEP \textbf{09}, 034 (2020)
doi:10.1007/JHEP09(2020)034
[arXiv:2004.14428 [hep-th]].
 

\bibitem{Cotler:2018zff}
J.~Cotler and K.~Jensen,
``A theory of reparameterizations for AdS$_3$ gravity,''
JHEP \textbf{02} (2019), 079
doi:10.1007/JHEP02(2019)079
[arXiv:1808.03263 [hep-th]].

\bibitem{Cotler:2020ugk}
J.~Cotler and K.~Jensen,
``AdS$_3$ gravity and random CFT,''
[arXiv:2006.08648 [hep-th]].
 
\bibitem{Afkhami-Jeddi:2020ezh}
N.~Afkhami-Jeddi, H.~Cohn, T.~Hartman and A.~Tajdini,
``Free partition functions and an averaged holographic duality,''
[arXiv:2006.04839 [hep-th]].

\bibitem{Maloney:2020nni}
A.~Maloney and E.~Witten,
``Averaging over Narain moduli space,''
JHEP \textbf{10}, 187 (2020)
doi:10.1007/JHEP10(2020)187
[arXiv:2006.04855 [hep-th]].
   
\bibitem{Maxfield:2019hdt}
H.~Maxfield,
``Quantum corrections to the BTZ black hole extremality bound from the conformal bootstrap,''
JHEP \textbf{12}, 003 (2019)
doi:10.1007/JHEP12(2019)003
[arXiv:1906.04416 [hep-th]].  
   
\bibitem{Maxfield:2020ale}
H.~Maxfield and G.~J.~Turiaci,
``The path integral of 3D gravity near extremality; or, JT gravity with defects as a matrix integral,''
[arXiv:2006.11317 [hep-th]].
   
\bibitem{Lunin:2002bj}
O.~Lunin, S.~D.~Mathur and A.~Saxena,
``What is the gravity dual of a chiral primary?,''
Nucl. Phys. B \textbf{655}, 185-217 (2003)
doi:10.1016/S0550-3213(03)00081-6
[arXiv:hep-th/0211292 [hep-th]].    
 
\bibitem{Eberhardt:2020bgq}
L.~Eberhardt,
``Partition functions of the tensionless string,''
[arXiv:2008.07533 [hep-th]].  
   
\bibitem{Giombi:2008vd}
S.~Giombi, A.~Maloney and X.~Yin,
``One-loop Partition Functions of 3D Gravity,''
JHEP \textbf{08} (2008), 007
doi:10.1088/1126-6708/2008/08/007
[arXiv:0804.1773 [hep-th]].


\bibitem{David:2009xg}
J.~R.~David, M.~R.~Gaberdiel and R.~Gopakumar,
``The Heat Kernel on AdS(3) and its Applications,''
JHEP \textbf{04} (2010), 125
doi:10.1007/JHEP04(2010)125
[arXiv:0911.5085 [hep-th]].

\bibitem{Castro:2011iw}
A.~Castro, R.~Gopakumar, M.~Gutperle and J.~Raeymaekers,
``Conical Defects in Higher Spin Theories,''
JHEP \textbf{02} (2012), 096
doi:10.1007/JHEP02(2012)096
[arXiv:1111.3381 [hep-th]].


\bibitem{Kaidi:2020ecu}
J.~Kaidi and E.~Perlmutter,
``Discreteness and Integrality in Conformal Field Theory,''
[arXiv:2008.02190 [hep-th]].


\bibitem{Vasiliev:1986qx}
M.~A.~Vasiliev,
``Extended Higher Spin Superalgebras and Their Realizations in Terms of Quantum Operators,''
 Fortsch. Phys. \textbf{36} (1988), 33--62.
 
 
\bibitem{Romans:1991wi}
L.~J.~Romans,
``The N=2 superW(3) algebra,''
 Nucl. Phys. B \textbf{369}, 403--432 (1992)
 doi:10.1016/0550-3213(92)90392-O
 
 
\bibitem{Banados:2015tft}
M.~Banados, A.~Castro, A.~Faraggi and J.~I.~Jottar,
``Extremal Higher Spin Black Holes,''
JHEP \textbf{04}, 077 (2016)
doi:10.1007/JHEP04(2016)077
[arXiv:1512.00073 [hep-th]].
 
 
\bibitem{Creutzig:2011fe}
T.~Creutzig, Y.~Hikida and P.~B.~Ronne,
``Higher spin AdS$_3$ supergravity and its dual CFT,''
JHEP \textbf{02} (2012), 109
doi:10.1007/JHEP02(2012)109
[arXiv:1111.2139 [hep-th]].

\end{thebibliography}
\end{document}